\documentclass[11pt,a4paper]{article}
\usepackage[utf8]{inputenc}
\usepackage{subcaption}
\usepackage{mathtools}
\usepackage{appendix}
\usepackage{jcappub}
\usepackage{amsmath}
\usepackage{slashed}

\author[a]{Chad Briddon,}
\author[a]{Clare Burrage,}
\author[a]{Adam Moss,}
\author[b]{and Andrius Tamosiunas}

\affiliation[a]{School of Physics and Astronomy, University of Nottingham, Nottingham, NG7 2RD,\\ United Kingdom}
\affiliation[b]{CERCA/ISO, Department of Physics, Case Western Reserve University, 10900 Euclid Avenue, Cleveland, OH 44106, USA}

\emailAdd{chad.briddon@nottingham.ac.uk}
\emailAdd{clare.burrage@nottingham.ac.uk}
\emailAdd{adam.moss@nottingham.ac.uk}
\emailAdd{andrius.tamosiunas@case.edu}

\arxivnumber{}

\title{Using machine learning to optimise chameleon fifth force experiments}

\abstract{The chameleon is a theorised scalar field that couples to matter and possess a screening mechanism, which weakens observational constraints from experiments performed in regions of higher matter density. One consequence of this screening mechanism is that the force induced by the field is dependent on the shape of the source mass (a property that distinguishes it from gravity). Therefore an optimal shape must exist for which the chameleon force is maximised. Such a shape would allow experiments to improve their sensitivity by simply changing the shape of the source mass. In this work we use a combination of genetic algorithms and the chameleon solving software SELCIE to find shapes that optimise the force at a single point in an idealised experimental environment. We note that the method we used is easily customised, and so could be used to optimise a more realistic experiment involving particle trajectories or the force acting on an extended body. We find the shapes outputted by the genetic algorithm possess common characteristics, such as a preference for smaller source masses, and that the largest fifth forces are produced by small `umbrella'-like shapes with a thickness such that the source is unscreened but the field reaches its minimum inside the source. This remains the optimal shape even as we change the chameleon potential, and the distance from the source, and across a wide range of chameleon parameters. We find that by optimising the shape in this way the fifth force can be increased by $2.45$ times when compared to a sphere, centred at the origin, of the same volume and mass.}

\begin{document}
\maketitle
\section{Introduction}
\label{sec:Introduction}

Through the use of a conformal transformation of the metric tensor, some modifications to the Einstein-Hilbert action can be recast as Einstein's general relativity (GR) but with a new scalar degree of freedom that couples to matter \cite{fujii2003scalar}. Such a scalar field could be the source of the `dark energy' we observe through the current expansion rate of the universe \cite{Copeland:2006wr}. This field would also mediate a new fundamental force due to its coupling to matter. This `fifth force' would be observed as deviations in gravitational phenomena from those predicted by GR. However, to date, all experimental results have been in agreement with the predictions of GR, placing strong constraints on the couplings between such a scalar field and matter, in models with a linear equation of motion \cite{Adelberger:2003zx, Kapner2007, adelberger2009torsion}. Consequently these fields are no longer viable dark energy candidates. This has motivated research into nonlinear scalar field models that possess screening mechanisms. These models have fifth forces that are suppressed in and around regions of relatively high density, resulting in weaker experimental constraints. Some examples include the symmetron, chameleon, and Galileon models \cite{Khoury:2003rn, Hinterbichler:2010es, Nicolis:2008in}.

\begin{figure}
    \centering
    \includegraphics{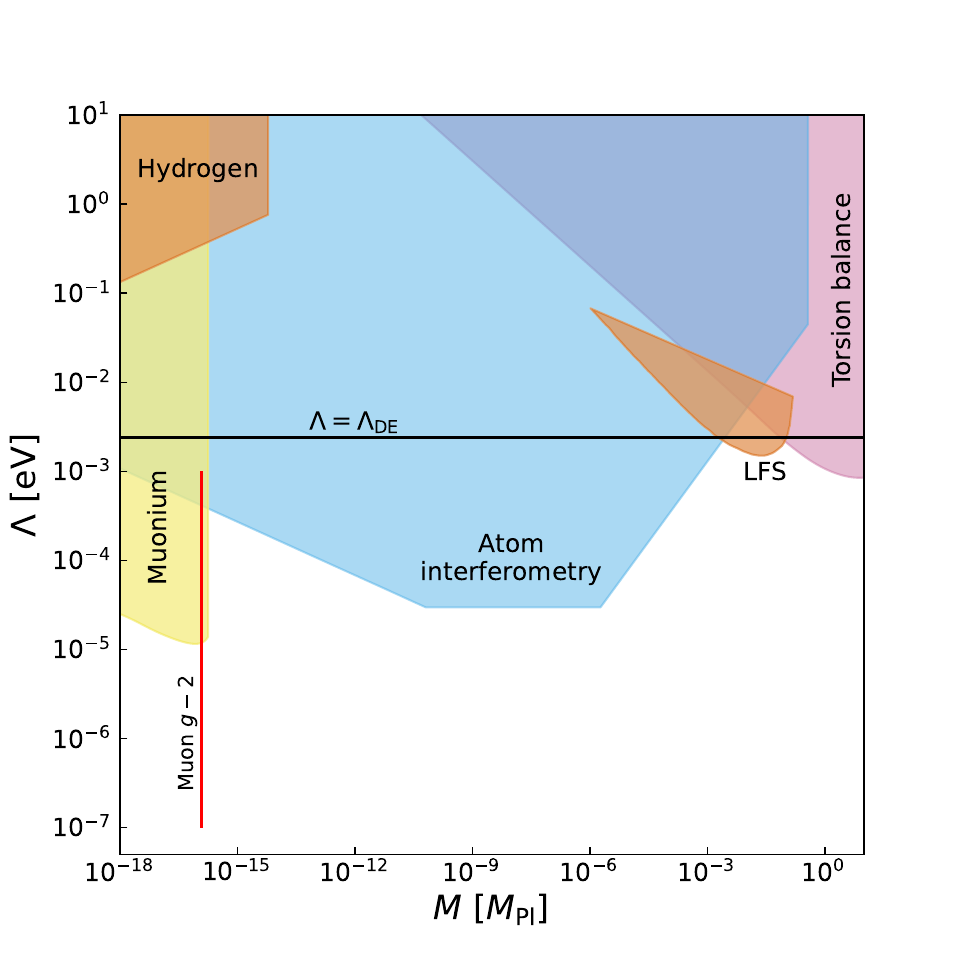}
    \caption{The current constraints (shaded regions) on the parameter space of the chameleon model with a $V(\phi)=\Lambda^5/\phi$ potential. $M$ controls the strength of the coupling to matter and $\Lambda$ the strength of the self-interactions of the scalar field. The black line indicates where $\Lambda$ is equal to the dark energy value, $\Lambda_{\rm DE}=2.4 \mbox{  meV}$. The red line indicates where the chameleon can explain the difference in the measured and theoretical values of the muon $g-2$ value \cite{Brax:2021owd}. LFS stands for Levitated Force Sensor. Figure reproduced with permission from Ref.~\cite{brax2023casimir}.}
    \label{fig:Chameleon parameter space}
\end{figure}

In this work we will focus on the chameleon, which is a class of models where the field has an effective mass that scales with the local matter-energy density \cite{Khoury:2003rn}. Consequently, in regions of high density, such as close to the Earth, the field's mass is larger and interactions are suppressed. Despite this, numerous experiments, including torsion balance \cite{Upadhye:2012qu}, Casimir \cite{Brax:2007vm, Elder:2019yyp, Sedmik:2021iaw, brax2023casimir}, levitated force sensors \cite{Yin:2022geb}, atom interferometry \cite{Burrage:2014oza, Hamilton:2015zga, Elder:2016yxm, Jaffe:2016fsh, Sabulsky_2019}, atomic spectroscopy \cite{Brax:2010gp, Brax:2022olf}, neutron bouncing \cite{Brax:2011hb, Ivanov:2012cb, Brax:2013cfa, Jenke:2014yel, Cronenberg:2015bol}, and neutron interferometry \cite{Lemmel:2015kwa, Li:2016tux}, have been able to constrain the chameleon parameter space as shown in Figure \ref{fig:Chameleon parameter space}. For a comprehensive review of the constraints on the chameleon model see Refs.~\cite{Brax:2021wcv, Burrage2018, Burrage:2016bwy}.

The nonlinearity of the chameleon's equations of motion makes calculating analytic solutions very difficult. Examples do exist in the literature but these are only for highly symmetrical source shapes and often involve approximations that limit the solution to specific regions of parameter space \cite{Khoury:2003rn, Silvestri2011, Burrage_2015}. Constraints from experiments and observations will either come from one of these analytic approximations or from numerically solving the equations of motion for the system being studied. The question may then arise: is the experimental system optimal for detecting chameleons?

We know that the approximate analytic field profiles around spherical and cylindrical sources, as expressed in \cite{Khoury:2003rn} and \cite{Burrage_2015} respectively, have different forms. Consequently, for sources of equal density and radial size, the resulting fifth force induced by the scalar field will differ even at the same radial distance from the source\footnote{For a comparison between the approximate analytic chameleon solutions around a spherical and cylindrical source see Figure $7$ in Ref.~\cite{Briddon_2021}.}. This means that in an experiment measuring fifth forces, choosing to use a spherical source over a cylinder will improve the experimental sensitivity. This creates an interesting contrast to Newtonian gravity where Gauss' law requires that the force experienced by a test particle is independent of the matter distribution contained in some exterior surface. Since a dependence on the source shape exists for the chameleon \cite{Burrage2018B, Burrage_2015,Jones-Smith:2011qnl, Pourhasan:2011sm}, there must exist a shape that maximises the force when taking all other parameters to be fixed. Using such an optimised source shape in current and near future experiments would yield improvements to experimental sensitivity and thereby strengthen the resulting model constraints. The aim of this work is to demonstrate a method by which this optimised source shape can be found. The exact shape will depend on what observable is being optimised and so will depend on the specific experiment.

In this work we will determine the rotationally symmetric shape which maximises the force at a single point a fixed distance from the source, inside a spherical vacuum chamber. We employ several parameterisations to mathematically describe our source shapes. This reduces the problem to finding the set of parameters that maximise the fifth force, although the dimension of this parameter space will inevitably be large if we wish it to encompass complex shapes. Using the software package SELCIE (\url{https://github.com/C-Briddon/SELCIE.git}) \cite{Briddon_2021}, meshes were generated of the source shape inside a spherical vacuum chamber, with an additional surface for measurements at a fixed distance from the source.\footnote{For a review of the challenges in numerically simulating screened theories in laboratory and astrophysical environments see Ref.~\cite{Vardanyan:2023jkm}, and for a related finite element code that can simulate the behavior of the chameleon in satellite experiments see Ref.~\cite{Levy:2022xni}.} SELCIE was also used to numerically solve for the chameleon field profile of the system and solve for the field gradient magnitude which is proportional to the force. We then iterate over the mesh vertices along the `measuring surface' and output the maximum value of the field gradient magnitude. The goal then is to find the set of parameters defining the source shape that maximise the field gradient magnitude outputted by this algorithm. To achieve this optimisation we used the DEAP software package (\url{https://github.com/deap/deap.git}) to automate the design and running of a `genetic algorithm' \cite{DEAP_JMLR2012}. These genetic algorithms are a class of machine learning algorithms that use concepts of Darwinian evolution to improve a population of solutions so that it converges to an extremum in the solution space \cite{koza2005genetic, Sloss2019}. They also do not require the problem being optimised to have an analytic form, and are efficient even for problems with a large solution space dimension.

We will begin this work by reviewing the chameleon model in section \ref{sec:Chameleon Model}. The details of the genetic algorithm will be explained in section \ref{sec:Genetic Algorithm}. In section \ref{sec:Shape Optimisation} we will define the shape parameterisations used and show the results of our investigation using the genetic algorithm when the measuring distance from the source and chameleon potential are fixed. Sections \ref{sec:Effects of Varying Measuring Distance} and \ref{sec:Effects of Varying n} will then explore the effects of changing the measuring distance and potential, respectively. We will then end with our conclusions from this work in section \ref{sec:Conclusion}.

\textbf{Conventions}: In this work calculations are performed in natural units ($c=1$ and $\hbar=1$), and we use the mostly plus sign convention for the metric, so that the Minkowski metric is $\eta_{\mu \nu} = {\rm diag}(-1,1,1,1)$. We also use the reduced Planck mass $M_{\rm pl}^2 = (8 \pi G_{\rm N})^{-1}$, where $G_{\rm N}$ is Newton's gravitational constant. Partial derivatives are expressed as $f_{,x} = \frac{\partial f}{\partial x}$ and $f_{,xy} = \frac{\partial^2 f}{\partial x \partial y}$.

\section{Chameleon model}
\label{sec:Chameleon Model}

The chameleon model \cite{Khoury:2003rn}, in the Einstein frame (with metric $g_{\mu \nu}$ and metric determinant $g$), has the action
\begin{equation}
\label{eq:Chameleon - Action}
    S = \int d^4x \sqrt{-g} \left( \frac{M_{\rm pl}^2}{2} R - \frac{1}{2} g^{\mu \nu} \nabla_{\mu} \phi \nabla_{\nu} \phi - V(\phi)\right) - \int d^4x \mathcal{L}_m(\varphi_m, \tilde{g}_{\mu \nu}),
\end{equation}
where $R$ is the Ricci scalar derived in the Einstein frame, and $V(\phi)$ is the potential of the chameleon scalar field $\phi$. The dynamics of other matter species, $\varphi_m$, are described by the Lagrangian $\mathcal{L}_m(\varphi_m, \tilde{g}_{\mu \nu})$, where $\tilde{g}_{\mu \nu}$ is the Jordan frame metric, which is related to the Einstein frame metric by the conformal transformation 
\begin{equation}
\label{eq:Jordan to Einstein frame}
    \Tilde{g}_{\mu \nu} = A^2(\phi) g_{\mu \nu}.
\end{equation}
In the action shown in equation (\ref{eq:Chameleon - Action}), we have assumed that the scalar field has a universal coupling to matter. Taking this coupling to be constant in $\phi$, the conformal factor is $A(\phi) = \exp{(\beta \phi/M_{\rm pl})}$, where $\beta$ is the dimensionless universal coupling constant \cite{Burrage2018}.

In this work we assume a field potential of the form
\begin{equation}
    \label{eq:Chameleon - Potencial}
    V(\phi) = \frac{\Lambda^{n+4}}{\phi^{n}},
\end{equation}
where $\Lambda$ is a mass scale and $n$ is an integer, following Ref.~\cite{Khoury:2003rn}. Application of the least action principle to equation (\ref{eq:Chameleon - Action}), and taking the matter component to be non-relativistic, the equation of motion of the chameleon field is
\begin{equation}
    \label{eq:Chameleon - EOM}
    \Box \phi = -\frac{n \Lambda^{n+4}}{\phi^{n+1}} + \frac{\beta \rho}{M_{\rm pl}},
\end{equation}
where $\Box$ is the d'Alembert operator, and $\rho$ is the matter-energy density. This equation resembles a Klein-Gordan equation, with the right-hand side being the derivative of the field's effective potential. Taking the left-hand side of equation (\ref{eq:Chameleon - EOM}) to be zero, we get that the field value that minimises the effective potential is
\begin{equation}
    \label{eq:Chameleon - Field minimum}
    \phi_{\rm min}(\rho) = \Lambda \left(\frac{n M_{\rm pl} \Lambda^{3}}{\beta \rho} \right)^{\frac{1}{n+1}}.
\end{equation}
The mass of the field at this minimum is then
\begin{equation}
    \label{eq:Chameleon - Compton wavelength}
    m^2_{\phi} = V''(\phi_{\rm min}) = n^{-\frac{1}{(n+1)}} (n+1) \Lambda^2 \left(\frac{\beta \rho}{M_{\rm pl} \Lambda^3}\right)^\frac{(n+2)}{(n+1)}.
\end{equation}
We see from equation (\ref{eq:Chameleon - Compton wavelength}) that the effective mass of the field (and by extension its Compton wavelength through the relation $\lambda = 1/m_{\phi}$) depends on $\rho$, and in such a way that as $\rho$ increases so does the mass. This results in the field becoming suppressed in regions of higher density. This is the chameleon’s screening mechanism which allows it to avoid constraints from Earth bound and solar system experiments while having a non-negligible contribution on cosmological scales.

In this investigation we solved for the chameleon field profiles around source masses of arbitrary shapes. To accomplish this, we used the software package SELCIE \cite{Briddon_2021} to numerically solve a static dimensionless version of equation (\ref{eq:Chameleon - EOM}), for the various shapes, using a finite element approach. To derive this dimensionless form we rescale the density and field values as $\hat{\rho} = \rho/\rho_0$ and $\hat{\phi} = \phi/\phi_{0}$, where $\rho_0$ is the vacuum density and $\phi_{0}$ is simply equation (\ref{eq:Chameleon - Field minimum}) evaluated at $\rho = \rho_0$. We will also work using lengths that are rescaled by the size of our domain, $L$, and can therefore rewrite the Laplacian as $L \nabla = \hat{\nabla}$. The dimensionless chameleon equation then has the form
\begin{equation}
\label{eq:Chameleon – dimensionless EOM}
	\alpha \hat{\nabla}^2 \hat{\phi} = -\hat{\phi}^{-(n+1)} + \hat{\rho},
\end{equation}
where the dimensionless constant $\alpha$ is defined as
 \begin{equation}
\label{eq:alpha_def}
    \alpha \equiv \left( \frac{M_{\rm pl} \Lambda}{L^2 \beta \rho_{0}}\right) \left( \frac{n M_{\rm pl} \Lambda^3}{\beta \rho_{0}} \right)^{\frac{1}{n+1}}.
\end{equation}

The force experienced by an unscreened test particle, with mass $m$, due to the chameleon field is 
\begin{equation}
\label{eq:Chamelon - Force}
    \vec{F}_{\phi} = - \frac{m \beta}{M_{\rm pl}}\vec{\nabla}\phi.
\end{equation}
Using the rescaling convention above, this formula can be written as
\begin{equation}
\label{eq:Chameleon – Force rescaled}
    \frac{|\vec{F}_{\phi}|}{|\vec{F}_{\rm g}|} = \left(\frac{\beta \phi_0}{L M_{\rm pl} |\vec{a}_{\rm g}|} \right) |\hat{\nabla} \hat{\phi}|,
\end{equation}
where we have also rescaled the force by the magnitude of the gravitational force $\vec{F}_{\rm g}$ which is related to the gravitational acceleration by $\vec{F}_{\rm g} = m \vec{a}_{\rm g}$. Therefore, for fixed model parameters, by maximising the rescaled field gradient magnitude, $|\hat{\nabla} \hat{\phi}|$, we will also be maximising the corresponding fifth force.

As shown in Ref.~\cite{Briddon_2021}, the rescaled Compton wavelength of the field at the minimum of the effective potential and when $\rho = \rho_0$ is 
\begin{equation}
    \label{eq:Chameleon - Compton wavelngth (Rescaled)}
    \hat{\lambda}_0^2 = \frac{\alpha}{(n+1)}. 
\end{equation}
Therefore when $\alpha \gg (n+1)$ the Compton wavelength will be much greater than unity. Physically this means the field does not have enough space in the unit sized chamber to reach its vacuum value. Consequently, the rescaled effective potential of the field (see right-hand side of equation (\ref{eq:Chameleon – dimensionless EOM})) can be approximated by $\hat{V}_{eff} \approx -\hat{\phi}^{-(n+1)}$ inside the vacuum region, since $\hat{\phi} \ll 1$ and $\hat{\rho} = 1$. If we then apply the substitution $\hat{\phi} = \alpha^{-1/(n+2)} \hat{\varphi}$, the resulting equation of motion for $\hat{\varphi}$ will be independent of $\alpha$ inside the vacuum region to leading order. The significance of this result is that by solving for $|\hat{\nabla} \hat{\phi}|$ for one value of $\alpha$, we can then determine the fifth force strength for all values where $\alpha \gg (n+1)$, by applying an overall rescaling to the solution. This statement remains true while the dense regions of the system are much more dense than the vacuum, because the value of $\varphi$ inside the dense regions will differ for different values of $\alpha$, but for sufficiently large density the differences become negligible as $\varphi \approx 0$. Therefore, assuming the above conditions to be satisfied we can express equation (\ref{eq:Chameleon – Force rescaled}) as
\begin{equation}
\label{eq:Chameleon – Force rescaled varphi}
	\frac{|\vec{F}_{\hat{\varphi}}|}{|\vec{F}_{g}|} = \frac{\beta \Lambda}{M_{pl} L |\vec{a}_g|} \left(n L^2 \Lambda^2 \right)^{\frac{1}{n+2}} |\hat{\nabla} \hat{\varphi}|,
\end{equation}
where $|\hat{\nabla} \hat{\varphi}|$ (which in the rest of this work will be referred to as the force) depends on $n$ and the profile of $\hat{\rho}$ but not on $\Lambda$, $\beta$ or $L$. This equation is also entirely independent of both the source and vacuum density values, so long as there exists a sufficiently large hierarchy between them. We also see from this equation the importance of the domain size on the measured fifth force. Since $\beta$ and $\Lambda$ are constants, and $\hat{\varphi}$ is independent of $L$ (as long as $\alpha \gg (n+1)$), the total dependence of the domain size, $L$, on the above expression is $L^{-n/(n+2)}$. This means smaller systems will yield higher sensitivity to fifth forces. For the case where $n=1$ equation (\ref{eq:Chameleon – Force rescaled varphi}) can be used to express the acceleration acting on the test mass due to the scalar field in physical units as
\begin{equation}
    \label{eq:Chameleon – Force rescaled varphi (w units)}
    \left(\frac{|\vec{a}_{\hat{\varphi}}|}{\rm nm/s^2}\right) = 0.373 \beta \left(\frac{\Lambda}{\rm meV}\right)^{\frac{5}{3}} \left(\frac{L}{\rm cm}\right)^{-\frac{1}{3}} |\hat{\nabla} \hat{\varphi}|.
\end{equation}

\section{Genetic algorithm}
\label{sec:Genetic Algorithm}

In this work we use the genetic algorithm (GA) software DEAP \cite{DEAP_JMLR2012} to optimise the shape of a source mass such that the fifth force it gives rise to at a fixed distance is maximised. GAs are designed to take an initially random population of solutions to a problem and evolve them in order to find the optimum solution \cite{goldberg1989genetic, koza2005genetic}. This is done in three stages; selection, crossover, and mutation. In the first stage, selection, it is determined which individuals will survive to parent the next generation. To make this determination each solution in the population is evaluated using the function we wish to maximise (or minimise), which we will refer to as the fitness function. The fitness of an individual in our population is then the corresponding output of this fitness function, and will determine the likelihood of an individual surviving. As is the case with the later stages (crossover and mutation) there are many methods to determine survivors in our population. For a comparison of the different possible configurations of the GA see appendix \ref{apx:Genetic Algorithm Calibration}. The method we ultimately chose was a tournament method, where we randomly select $N_{\rm tournsize}$ individuals and return the one with the best fitness (largest value if searching for maximum and smallest value for minimum) \cite{Miller1995GeneticAT}. This process is repeated $N$ times, where $N$ is the size of our original population. Note that it is possible for individuals to be selected multiple times. This process imitates natural selection and results in a new population with an increased average fitness. It is these survivors that will be the parents of the next generation. The fact that an individual can be selected multiple times is analogous to members of a species that have more offspring, thereby making its characteristics more abundant in the next generation.

Reproduction is replicated in the crossover stage, whereby individuals in the population are paired and each pair has some probability, $P_x$, of reproducing. If this occurs then each attribute of the solution pair (e.g. their coefficients) have a $p_x$ chance of being swapped. The pair has now become a pair of children that replace the parents in the population. This process combined with the selection process allows individuals with better fitness values to share features with a larger number of other individuals at a faster rate than those with worse fitness. Overall, this will lead the population as a whole to flow towards optimum fitness.

A general problem with optimisation algorithms is that of local maxima. If an individual finds such a solution, it can lead the rest of the population to it without finding the global maximum (e.g., see Ref.~\cite{kochenderfer2019algorithms}). The strategy used by a GA to address this is contained in the mutation stage. This stage is inspired by the mutations in natural organisms required for evolution, and modifies the information present in the current population. In this stage, each individual in the population has some probability, $P_m$, of receiving a mutation. This involves changing some of the attributes of the individual through some process involving randomness. In this work, since our parameters are continuous, each coefficient of the mutated individual has probability $p_m$ of receiving a Gaussian perturbation with standard deviation $\sigma$. This allows individuals to escape from local maxima while introducing more variation into the population, allowing more of the parameter domain to be explored.

In total, this algorithm has 7 parameters that must be set by the user, $\{$$N$, $N_{\rm tournsize}$, $P_x$, $P_m$, $p_x$, $p_m$, $\sigma$$\}$. In this work $\{$$N$, $N_{\rm tournsize}$, $P_x$, $P_m$, $p_x$$\} = \{$$105$, $5$, $0.7$, $0.1$, $0.5$$\}$. The values of $\{p_m, \sigma\}$ will depend on the number and domain of the coefficients. The process by which we decided on the methods described above, and the values of the parameters, is discussed in Appendix \ref{apx:Genetic Algorithm Calibration}.

\section{Shape optimisation}
\label{sec:Shape Optimisation}

To begin finding the shape which optimises the chameleon force, we will need to first define a space of shapes through some parameterisation. We note that any parameterisation of shapes, for reasons of practicality, must be described by a finite number of coefficients. However, in a complete basis an infinite number of coefficients is required to reproduce all possible shapes. Therefore, any optimisation performed using a particular parameterisation will at best find the global maximum of a finite sub-space of shapes, and not of the infinite space of shapes. We therefore investigated multiple shape parameterisations to probe different regions of the shape space in the hope of discovering consistent solutions, or even shared features that point to a more globally optimum solution. For each of these parameterisations we will require a function that maps the parameters to the measurable of interest. We shall henceforth refer to this function as the fitness function, given we intend to maximise it using the GA discussed in section \ref{sec:Genetic Algorithm}. In this work, the measurable outputted by this fitness function is the force at a single point that lies somewhere on a measuring surface a distance $d$ from the source, which lies inside of a spherical vacuum chamber. We restricted our search to rotationally symmetric sources as it allowed us to represent the systems using 2D meshes by imposing the symmetry on the field solutions. The reason for this was to reduce the runtime and memory requirements, although in principle our approach can be extended to more arbitrary systems by using 3D meshes instead.

The fitness function starts by taking the shape parameters and from them derives a list where each element contains the coordinates of a point. These points connected in sequence by straight lines define the boundary of the source. The python package SELCIE is then used to construct a mesh representing the system. This includes the source, a measuring surface a distance $d$ from the source, the vacuum, and the chamber wall. The mesh settings used were such that the cell size was a minimum at the boundary of two regions, with a size of $\rm CellSizeMin = 10^{-4}$, grew linearly for a distance $\rm DistMax=0.1$, beyond which the cell size was $\rm CellSizeMax = 0.01$. The exception to this was the boundary between the vacuum and the chamber wall which had $\rm CellSizeMin=5 \times 10^{-4}$ and $\rm DistMax=0.5$ instead, since this improved the runtime while contributing negligibly to the solutions' accuracy. The thickness of the chamber wall was set to $0.05$. Using this mesh, SELCIE is then used to calculate the scalar field profile and the force induced by the field. Finally, the magnitude of the force is evaluated at the mesh vertices that form the measuring surface. The largest of these values is then the output of the fitness function. The optimal shape will correspond to the global maximum of this function on the shape parameter space. For shape spaces with a dimensionality of one or two, it is reasonable to perform an iterative search of the parameters. However, as the dimension of our parameterised shape space increases this method becomes impractical. We will therefore use the GA, as described in section \ref{sec:Genetic Algorithm}, to find the shape parameters that maximise the outputted force value.

In the following sections we will define our shape parameterisations and will probe the corresponding space of shapes to find the shapes that maximise the chameleon force. We will then compare these optimal shapes to determine if there are any consistent features among them and whether these are indicative of a superior class of shapes.

\subsection{Ellipsoids}
\label{sec:Ellipsoid}
\begin{figure}
    \centering
    \begin{subfigure}[b]{0.49\textwidth}
        \centering
         \includegraphics[width=\textwidth]{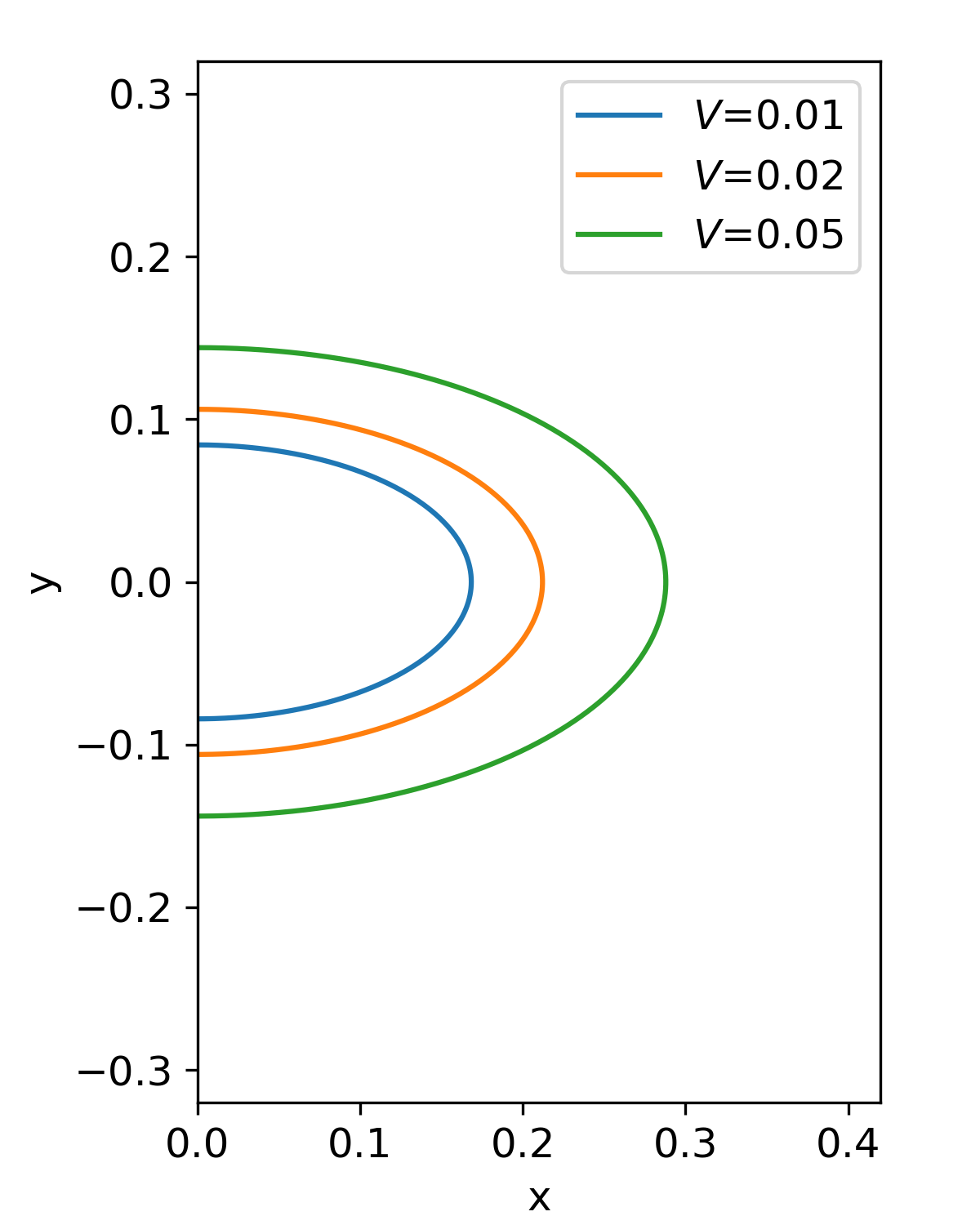}
         \label{fig:Ellipsoid - Example (1)}
    \end{subfigure}
    \hfill
    \begin{subfigure}[b]{0.49\textwidth}
        \centering
         \includegraphics[width=\textwidth]{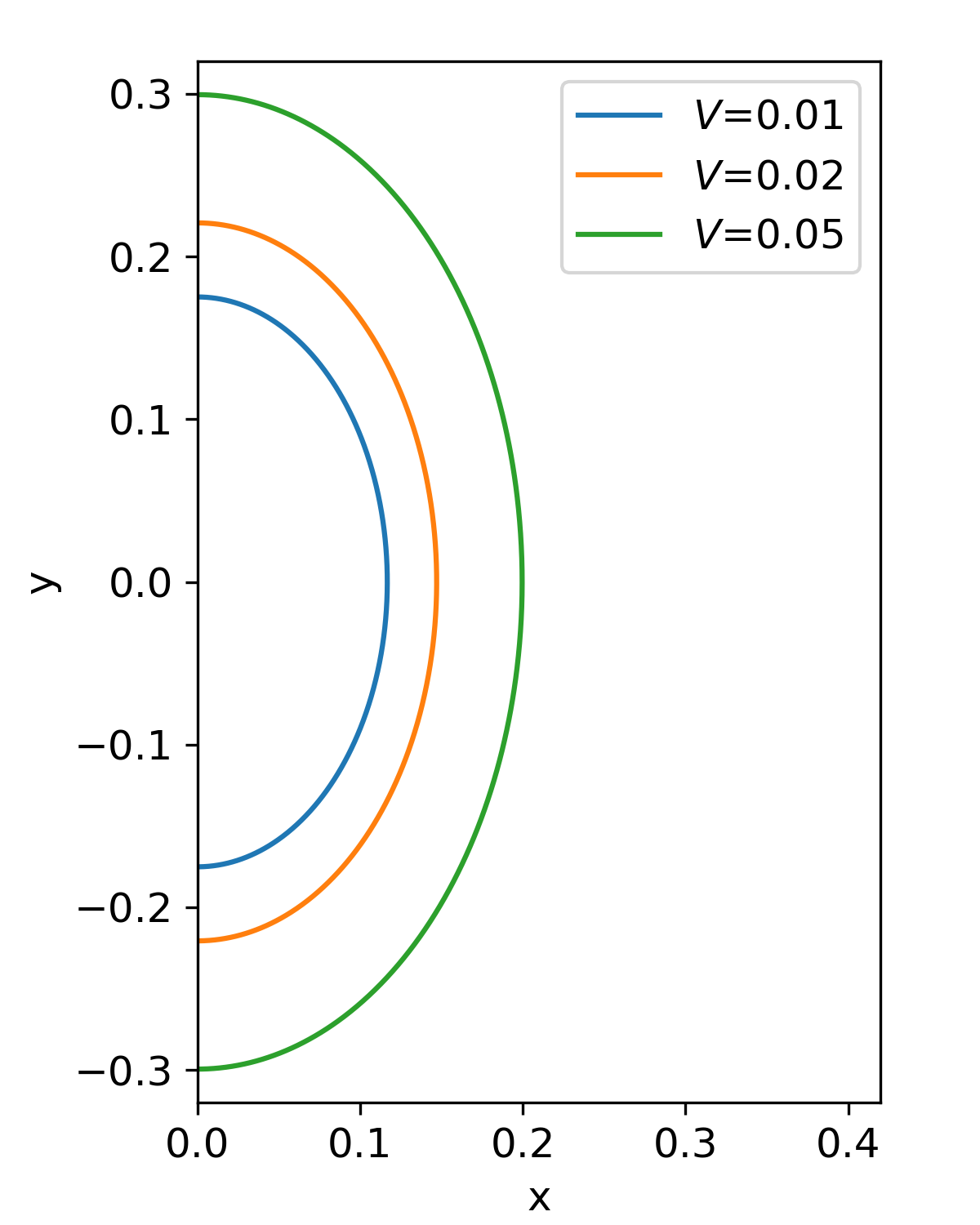}
         \label{fig:Ellipsoid - Example (2)}
    \end{subfigure}
    \caption{Boundary curves of elliptical shapes of varying volume and axis-ratio. The \textbf{left} plot contains ellipses with $\epsilon=0.5$, while the \textbf{right} contains ellipses with $\epsilon=1.5$.}
    \label{figs:Ellipsoid - Examples (all)}
\end{figure}

We will begin our investigation with ellipsoidal shapes (for earlier work see Ref.~ \cite{Burrage_2015}). Being the simplest deviation from spherically symmetric sources, these shapes are parameterised by the horizontal and vertical axis radii, $r_x$ and $r_y$ respectively. Since the shape space is only 2-dimensional, it can be thoroughly investigated without the need of the GA. The boundary of the ellipse in the $(x,y)$ plane is $(x, y) = (r_x \sin(\theta), r_y \cos(\theta))$, for $\theta = [0, 2 \pi)$. For our purposes, however, it will be more useful to parameterise the ellipsoids by their axis-ratio, $\epsilon=r_y/r_x$, and the volume of the ellipsoid
\begin{equation}
    \label{eq:Ellipsoid - volume}
    V = \frac{4 \pi}{3} r_x^2 r_y,
\end{equation}
where we have used the fact that our source has rotational symmetry around the $y$-axis. For such ellipsoids with a volume $V$ and axis ratio $\epsilon$, their axis radii can be expressed as $(r_x, r_y) = (\zeta, \zeta \epsilon)$, where the scale factor $\zeta$ is
\begin{equation}
    \label{eq:Ellipsoid - scale factor}
    \zeta = \left( \frac{3 V}{4 \pi \epsilon} \right)^{1/3}.
\end{equation}
Some examples of ellipsoids parameterised by $(V, \epsilon)$ are shown in Figure \ref{figs:Ellipsoid - Examples (all)}.

\begin{figure}
    \centering
    \begin{subfigure}[b]{0.49\textwidth}
        \centering
         \includegraphics[width=\textwidth]{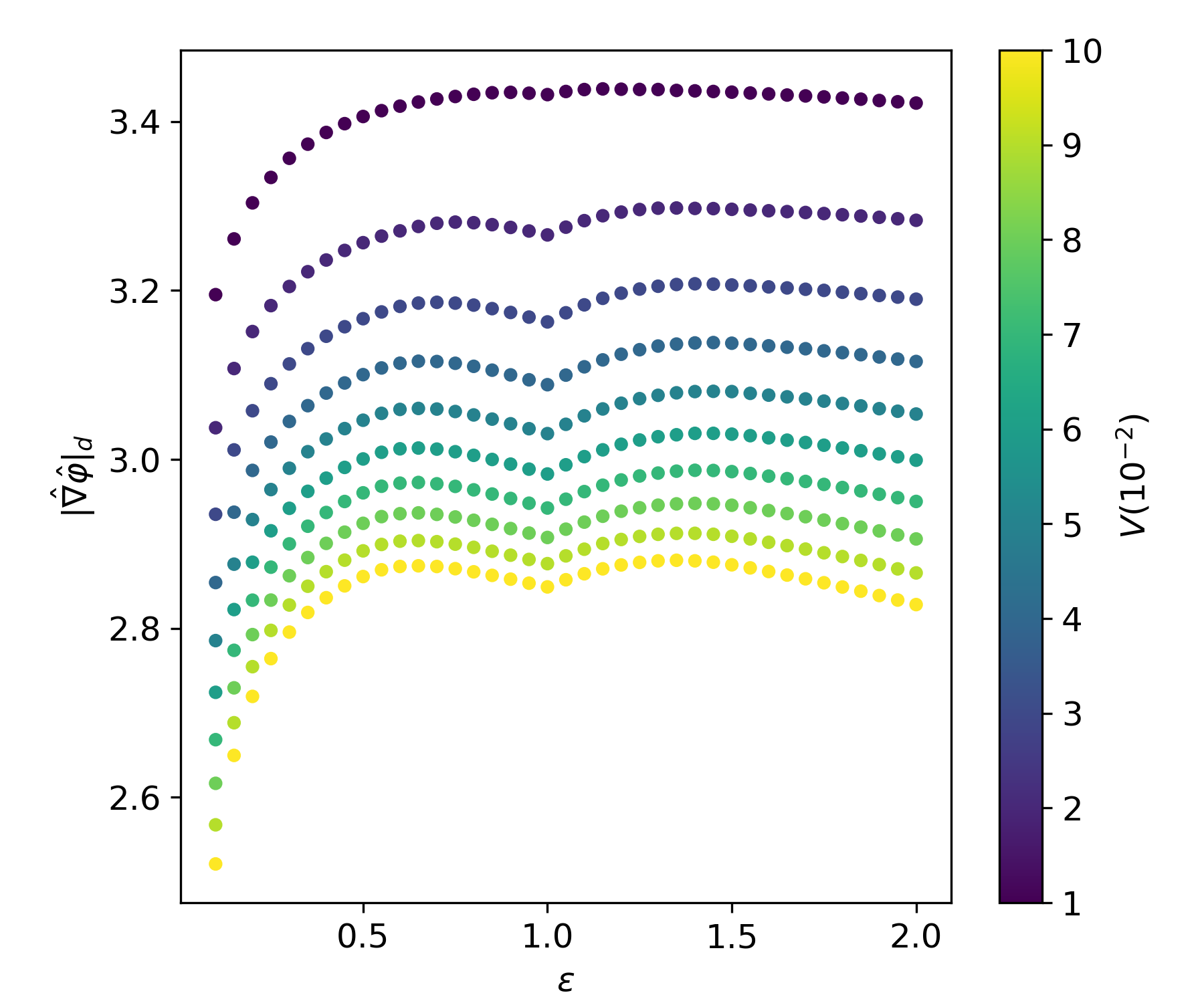}
         \caption{}
         \label{fig:Ellipsoid - Force_vs_ratio (1)}
    \end{subfigure}
    \hfill
    \begin{subfigure}[b]{0.49\textwidth}
        \centering
         \includegraphics[width=\textwidth]{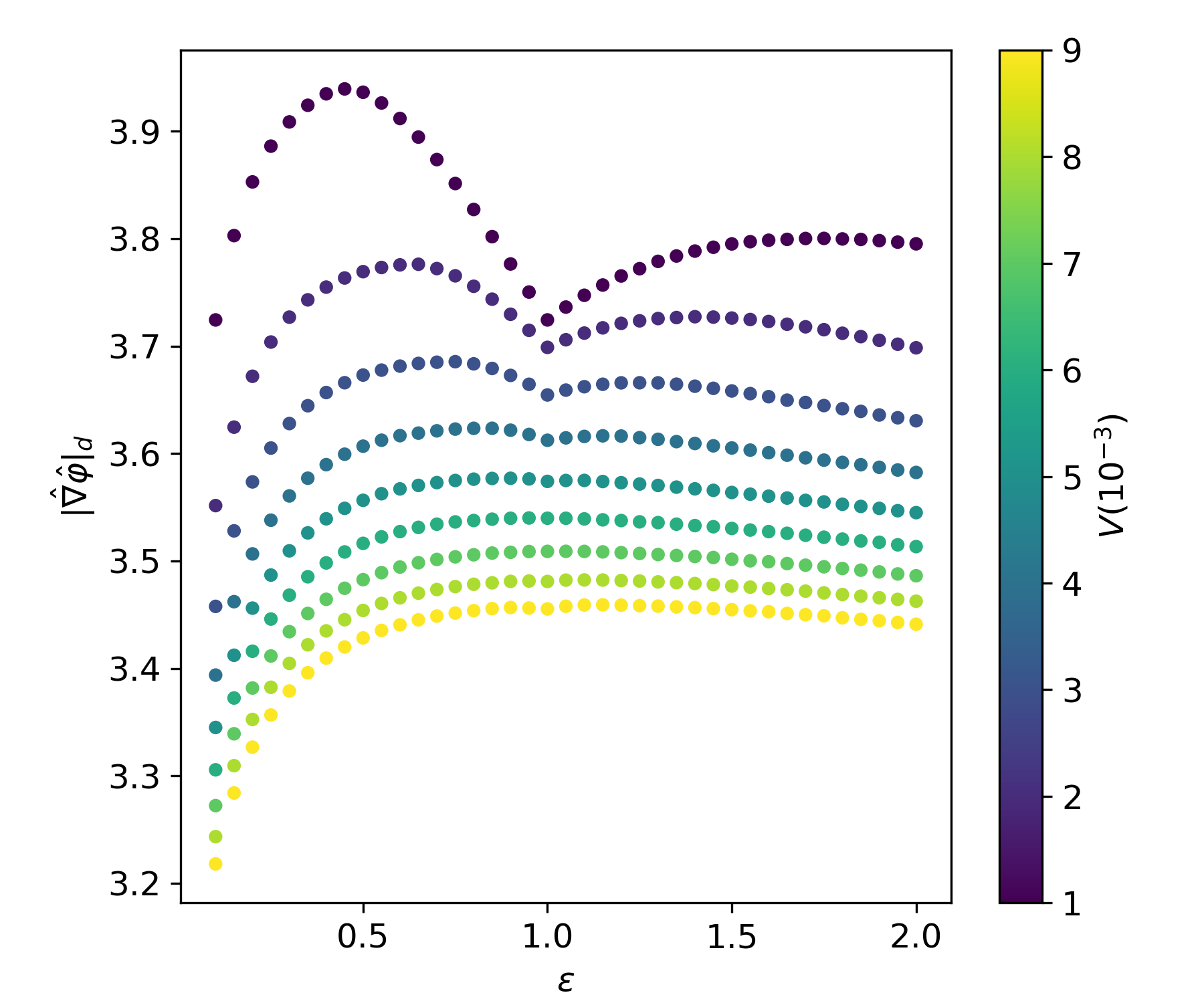}
         \caption{}
         \label{fig:Ellipsoid - Force_vs_ratio (2)}
    \end{subfigure}
    \hfill
    \begin{subfigure}[b]{0.49\textwidth}
        \centering
         \includegraphics[width=\textwidth]{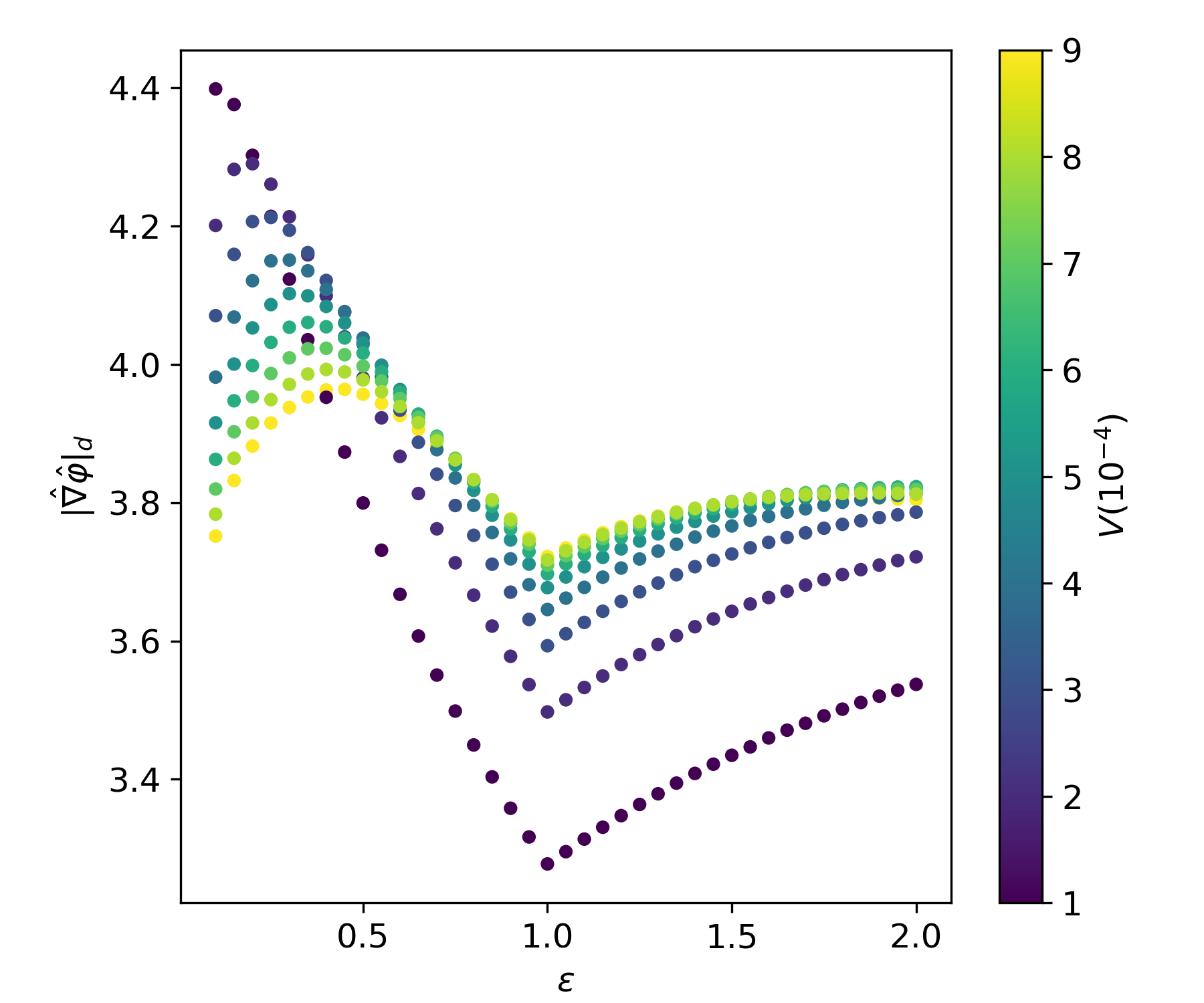}
         \caption{}
         \label{fig:Ellipsoid - Force_vs_ratio (3)}
    \end{subfigure}
    \hfill
        \begin{subfigure}[b]{0.49\textwidth}
        \centering
         \includegraphics[width=\textwidth]{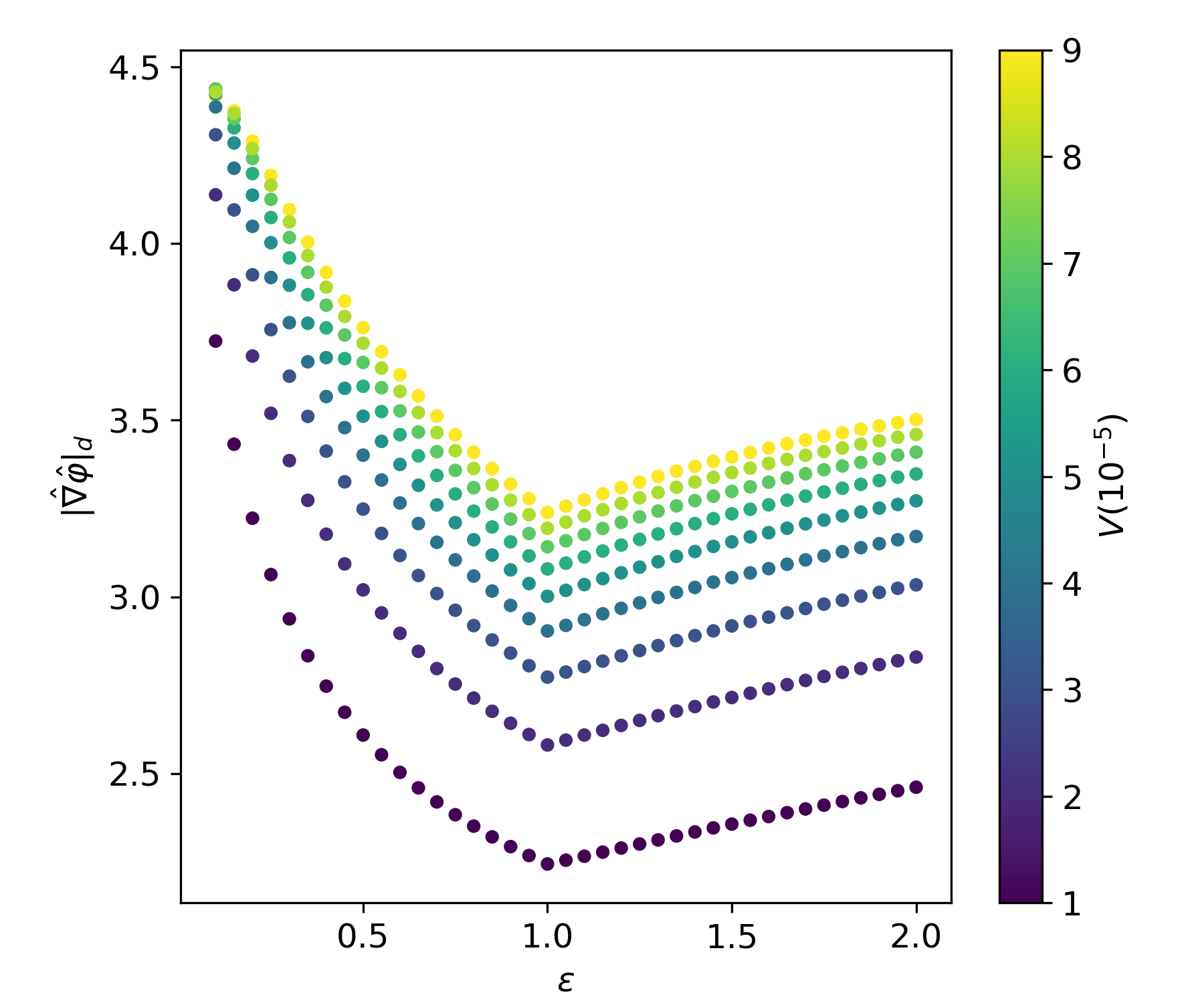}
         \caption{}
         \label{fig:Ellipsoid - Force_vs_ratio (4)}
    \end{subfigure}
    \caption{The measured maximum force (at distance 0.05 from the surface of an ellipsoid) against axis ratio $\epsilon=r_y/r_x$. The colour map represents the volume of the ellipsoid. Each subplot shows the results for a different order of magnitude in volume, with the largest volumes in the top left and the smallest in the bottom right.}
    \label{figs:Ellipsoid - Force_vs_ratio (all)}
\end{figure}

\begin{figure}
    \centering
    \includegraphics{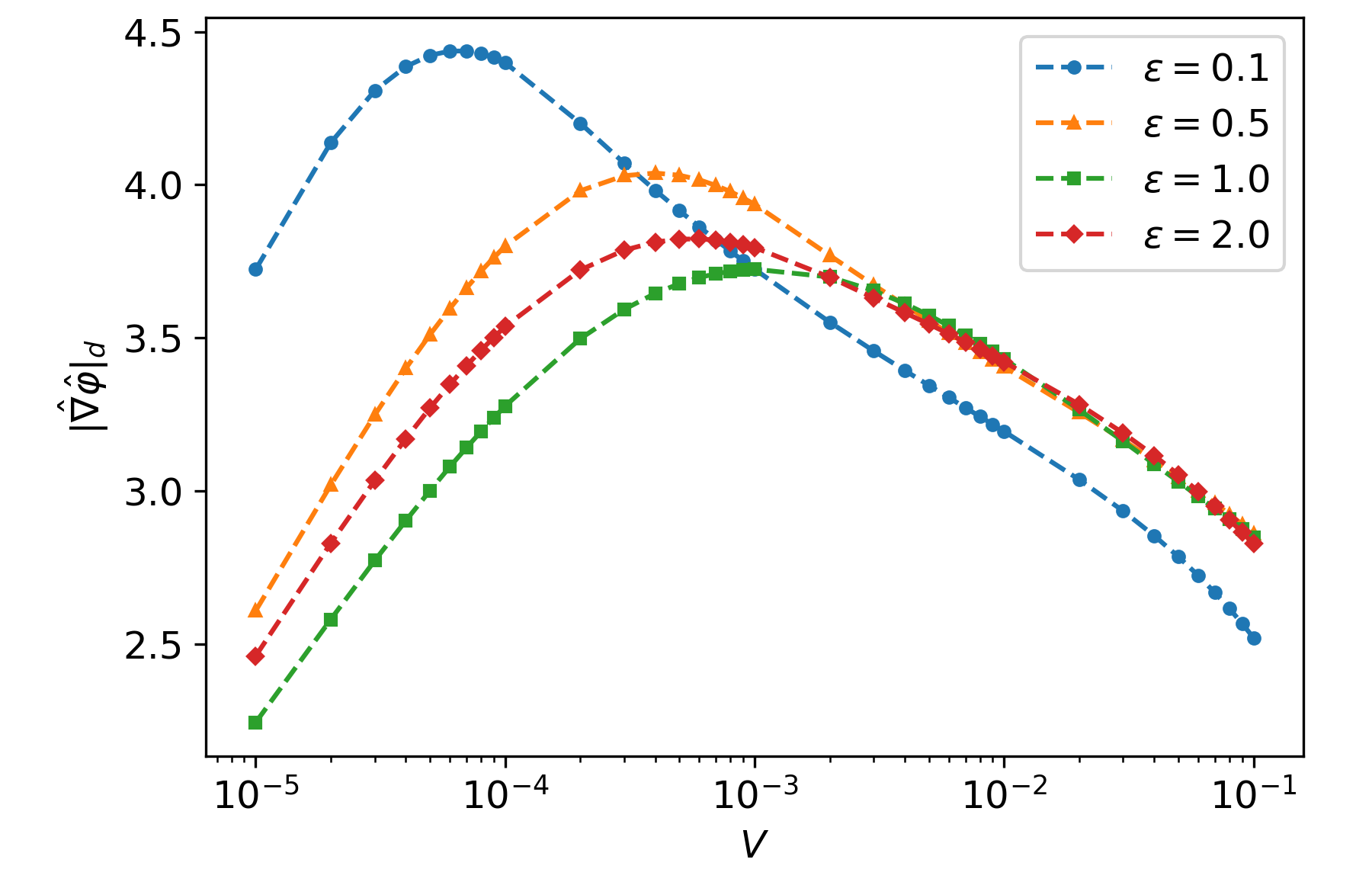}
    \caption{Measured maximum force (at distance 0.05 from the surface of the ellipsoid) against source volume, for a range of axis-ratio values.}
    \label{fig:Ellipsoid - Force_vs_volume}
\end{figure}

Figure \ref{figs:Ellipsoid - Force_vs_ratio (all)} shows the magnitude of the resulting force, at a distance $d=0.05$ from the source, for the values $\epsilon = [0.01, 2.0]$ and $V = [10^{-5}, 0.1]$. A note-worthy feature of this plot is that slight deviations from spherical sources appear to always result in an increase in the measured force, the location of which is around the pointed region of the ellipsoid as predicted in Ref.~\cite{Jones-Smith:2011qnl}. We also note that Figures \ref{fig:Ellipsoid - Force_vs_ratio (1)} and \ref{fig:Ellipsoid - Force_vs_ratio (2)} show that for fixed $\epsilon$ as $V$ decreases the force increases. However, Figures \ref{fig:Ellipsoid - Force_vs_ratio (3)} and \ref{fig:Ellipsoid - Force_vs_ratio (4)} show that this trend reverses after some critical volume $V_c(\epsilon)$. This can be seen clearly in Figure \ref{fig:Ellipsoid - Force_vs_volume} where we have plotted the force against volume for a range of $\epsilon$-values.

\begin{figure}
    \centering
    \includegraphics{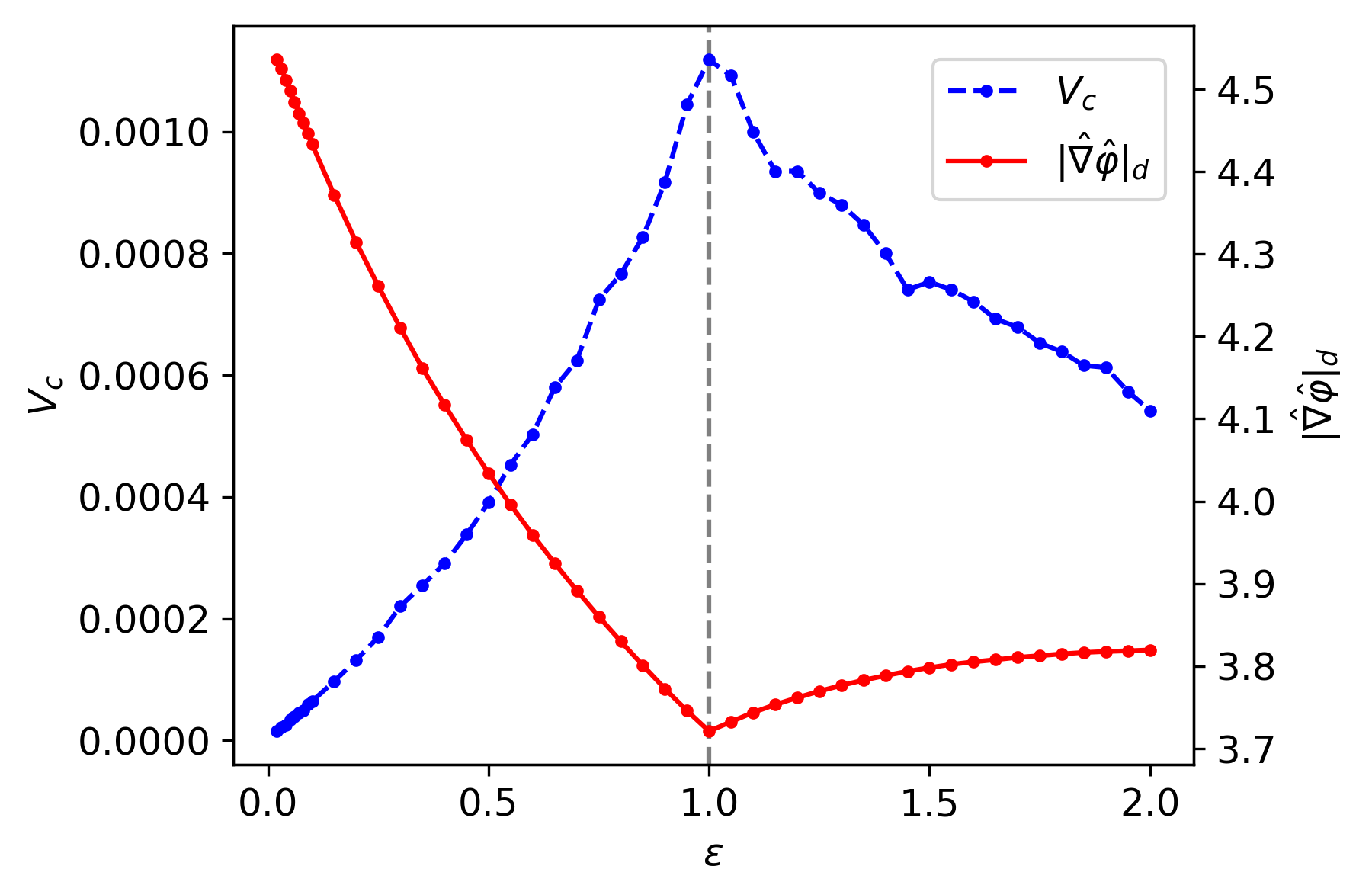}
    \caption{Critical volume (blue) and corresponding force measurement (red) against axis-ratio, $\epsilon$. The vertical gray dashed line at $\epsilon=1$ indicates the location of the spherical case.}
    \label{fig:Ellipsoid - Critical_volume_Force_vs_ratio}
\end{figure}

Figure \ref{fig:Ellipsoid - Critical_volume_Force_vs_ratio} shows this critical volume $V_c(\epsilon)$, at which the force is maximised, for a range of $\epsilon$. In this plot we also show the maximal force obtained for the corresponding ellipsoids. We acknowledge that the plot of $V_c(\epsilon)$ is not entirely smooth. This is because when performing our simulations we use a mesh of finite precision, resulting in a small numerical uncertainty. Although the relative uncertainty on the force is of the order $10^{-4}$, the finite mesh precision leads to a larger uncertainty on our measurement of $V_c$. Finally, we acknowledge that our plots would indicate that the ideal ellipsoid lies outside our parameter range. This is because of computational limitations of solving the field equation using meshes capable of resolving such small and elongated shapes. In practise, however, a very small ellipsoid will be impractical in realistic experiments.


\subsection{Legendre polynomial shapes}
\label{sec:Legendre polynomial shapes}

\begin{figure}
    \centering
    \includegraphics{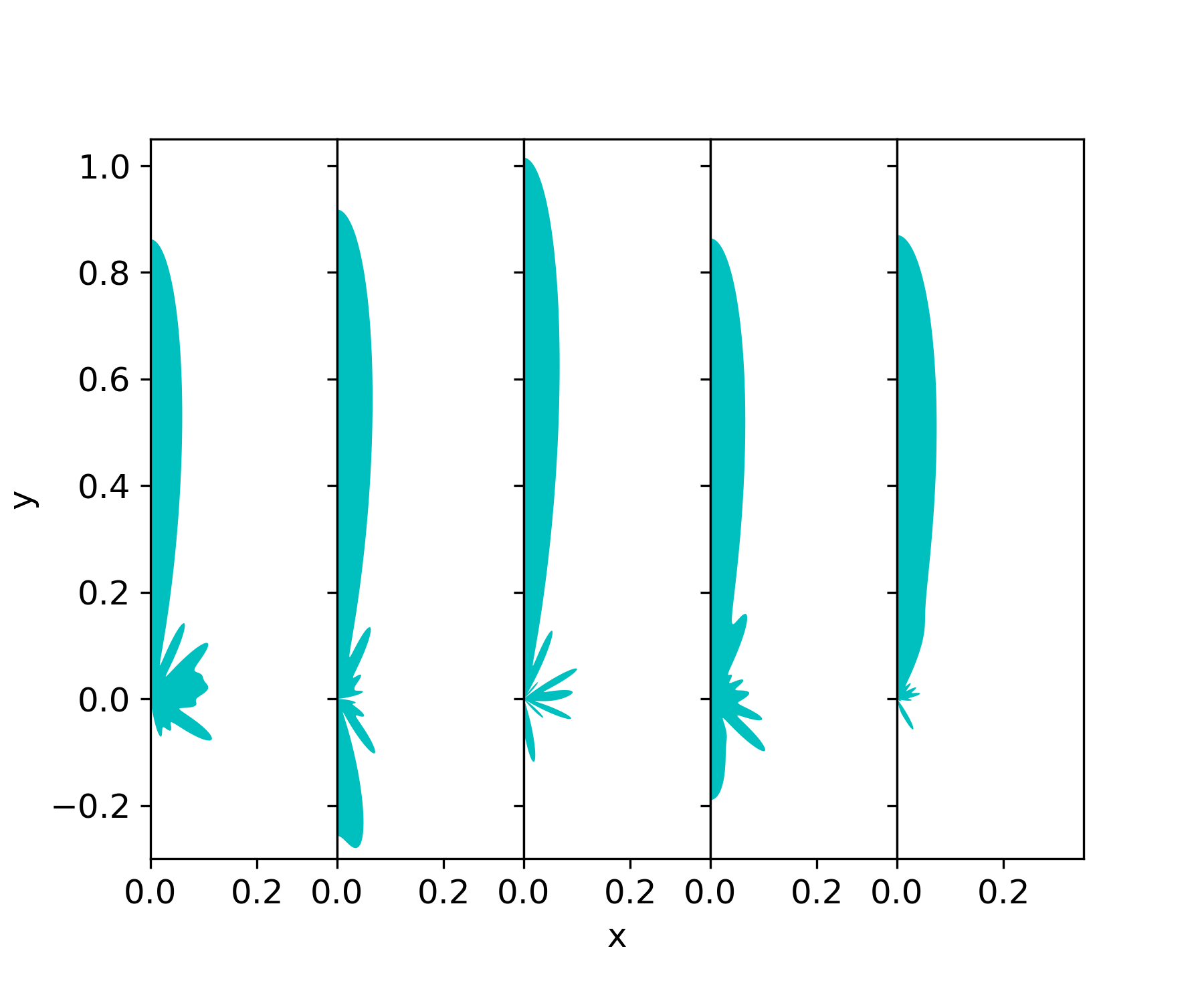}
    \caption{Examples of Legendre polynomial shapes defined using $N=20$ coefficients and with a volume fixed to $V=0.01$.}
    \label{fig:Legendre - Examples}
\end{figure}

Legendre polynomials, $P_n$, are a special set of polynomials that are solutions to the differential equation
\begin{equation}
    \label{eq:Legendre - definition}
    (1-x^2) P_{n}''(x) - 2x P_{n}'(x) + n(n+1) P_{n} = 0.
\end{equation}
This set of functions has the property
\begin{equation}
    \label{eq:Legendre - Basis equation}
    \int^{1}_{-1} P_n(x) P_m(x) dx = \frac{\delta_{nm}}{2n+1},
\end{equation}
showing that it forms a complete basis in the interval $[-1,1]$. This means if we define a set of curves in polar coordinates $(R(\theta), \theta)$ where
\begin{equation}
    \label{eq:Legendre - Curve equation}
    R(\theta) = \sum_{n=0}^N a_n P_n(\cos{\theta})
\end{equation}
and $a_n$ are the series coefficients, then in the limit $N \xrightarrow{} \infty$ the set will grow to contain every closed curve that is also symmetric around the y-axis. However, we must still consider how we relate these curves to shapes. For example, for some sets of $a_n$, it is possible for $R(\theta)$ to take negative values, which can result in the curve intersecting itself. In this work we take positive $r<R(\theta)$ to be inside the shape, and to avoid complications will set $R(\theta) = R_{\rm min}$ whenever the value of $R(\theta)$ given by equation (\ref{eq:Legendre - Curve equation}) is less than $R_{min}$. To investigate the shape dependence of the force, we keep the volume of the source fixed. We therefore define a mapping (see Appendix \ref{apx:Volume constraint}) from one set of coefficients, $a_n$, to a new set where the corresponding Legendre polynomial shape is similar to the original but with a volume $V$. Some examples of the kinds of shapes contained inside this class are shown in Figure \ref{fig:Legendre - Examples}.

It can be shown that the Legendre polynomials, in the domain $[-1, +1]$ have a maximum at $P(\theta=0)=1$. Constraining the coefficients $a_n$ to be positive, this means the maximum value taken by equation (\ref{eq:Legendre - Curve equation}) is
\begin{equation}
    \label{eq:Legendre R_max}
    R_{max} = \sum_i a_i.
\end{equation}
For practical reasons, we will only consider Legendre polynomial shapes that are contained entirely inside the vacuum chamber, e.g. $R_{max} < 1$ (after constraining the volume).

\begin{figure}
    \centering
    \includegraphics{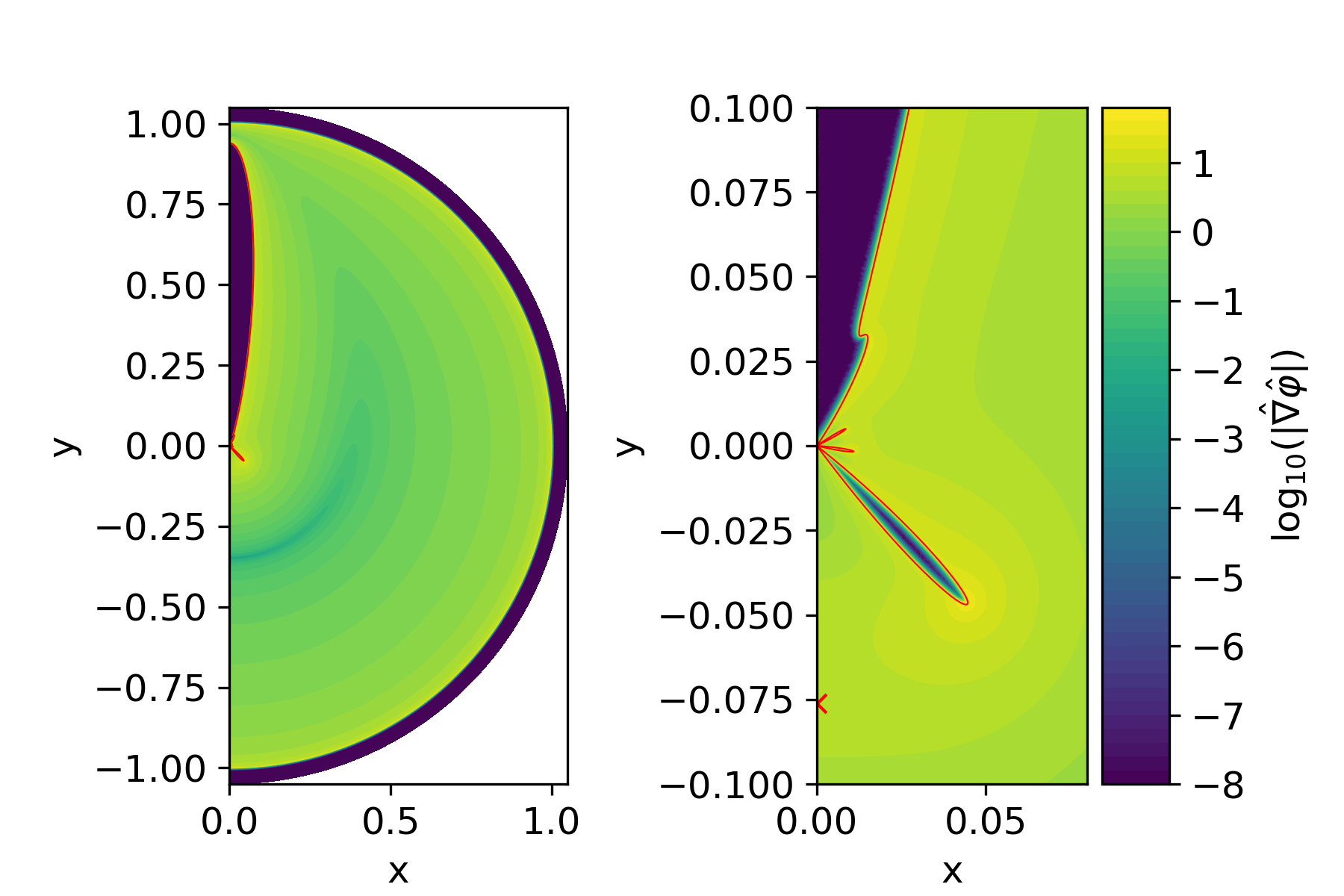}
    \caption{The chameleon force profile of an $N=20$ Legendre polynomial source (boundary of which is indicated by the red line) inside a spherical vacuum chamber. The \textbf{left} plot contains the full solution while the \textbf{right} is zoomed in on the central region. In the \textbf{right} plot the red cross on the y-axis below the source, at around $y=-0.075$, indicates the position where the force (at a distance from the source of $d=0.05$) is maximised. The value of this maximal force is $|\hat{\nabla} \hat{\varphi}|_{d} = 5.07$.}
    \label{fig:Legendre_GA_result_example}
\end{figure}

Using the method described in section \ref{sec:Shape Optimisation} we can obtain a force value from the set of coefficients $a_n$ that define a Legendre polynomial shape. The number of coefficients used was $N=20$; however, after applying the volume constraint the number of dimensions of the shape parameter space is reduced to $19$. Because of the large number of dimensions of the shape space we will use the GA discussed in section \ref{sec:Genetic Algorithm} to find the shape, parameterised by Legendre polynomials, that maximises the force at a distance $d=0.05$ from the source evaluated at a single point. For a volume of $V=0.01$ the shape outputted by the GA generated a force of $|\hat{\nabla} \hat{\varphi}|_{d} = 5.07$ (located along the y-axis), which is a $48\%$ increase when compared to a spherical source of the same volume centred at the origin. The profile of the scalar force induced by the source is shown in Figure \ref{fig:Legendre_GA_result_example} along with the position where the force is maximised.

\begin{figure}
    \centering
    \includegraphics{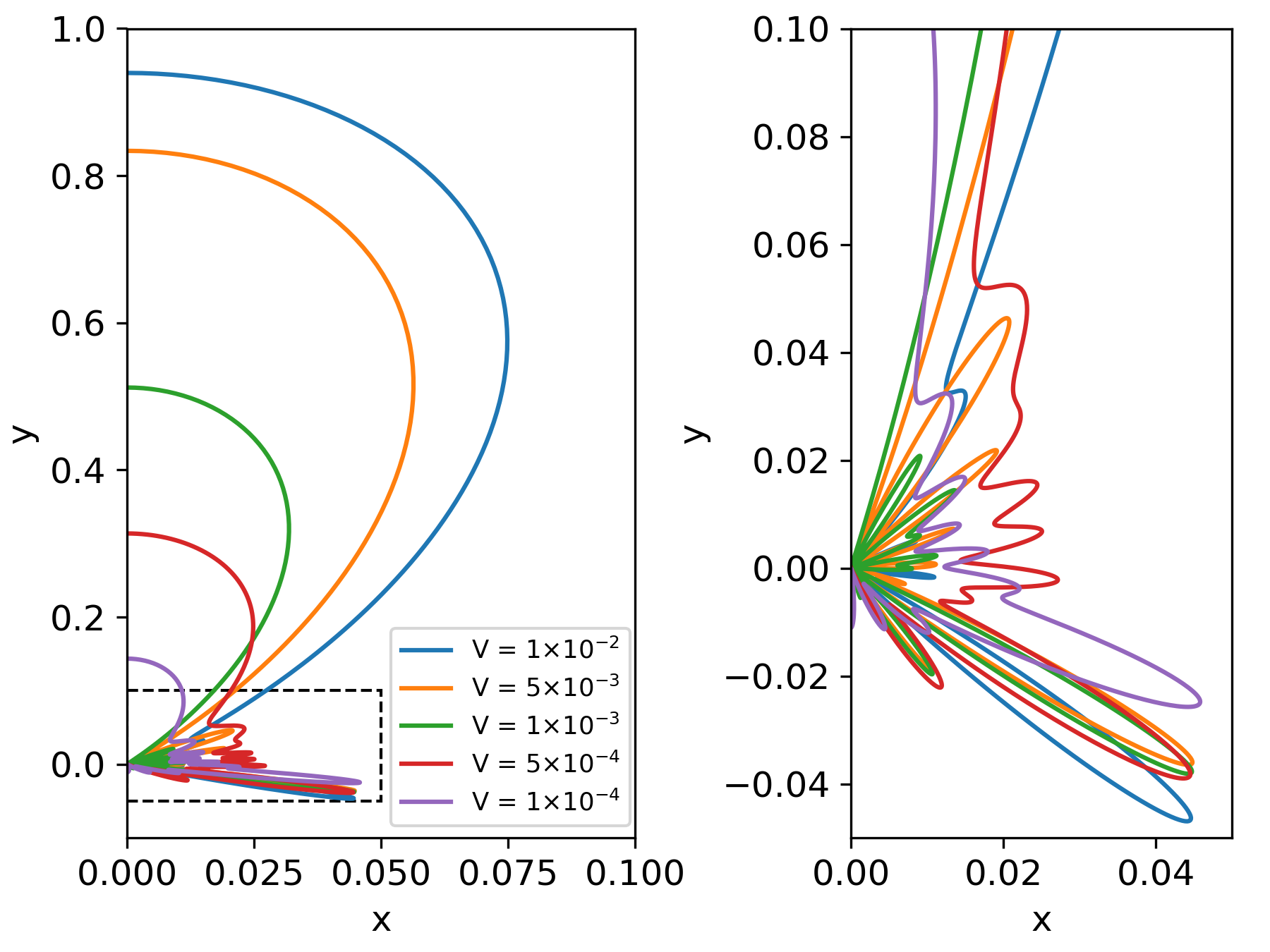}
    \caption{Shapes outputted by the GA when finding the optimal Legendre polynomial shape with $N = 20$ coefficients. Each line indicates a shape with a different volume, as shown in the legend, from $V=10^{-4}$ to $V=10^{-2}$. The \textbf{left} plot shows the full shapes, while the \textbf{right} is a zoomed in image of the area inside the dashed line.}
    \label{fig:Legendre - Compare R}
\end{figure}

We repeated our GA simulation for a range of other volumes, and found that as the volume is decreased the force becomes larger, as was the case with the ellipsoids. We also observed that the `umbrella'-like feature near $\theta = 3\pi/4$ in Figure \ref{fig:Legendre_GA_result_example} appears to be a common feature in each case, with the extra volume being diverted to the upper region, as seen in Figure \ref{fig:Legendre - Compare R}. Since our populations have independently evolved this feature, it suggests this may be the most important feature of the final shape and that it is independent of the total volume. This aspect will be further investigated in section \ref{sec:Is the Umbrella the Impactful Feature?}.

\subsection{Radially-parameterised shapes}
\label{sec:Radially-Parameterised Shapes}

\begin{figure}
    \centering
    \includegraphics{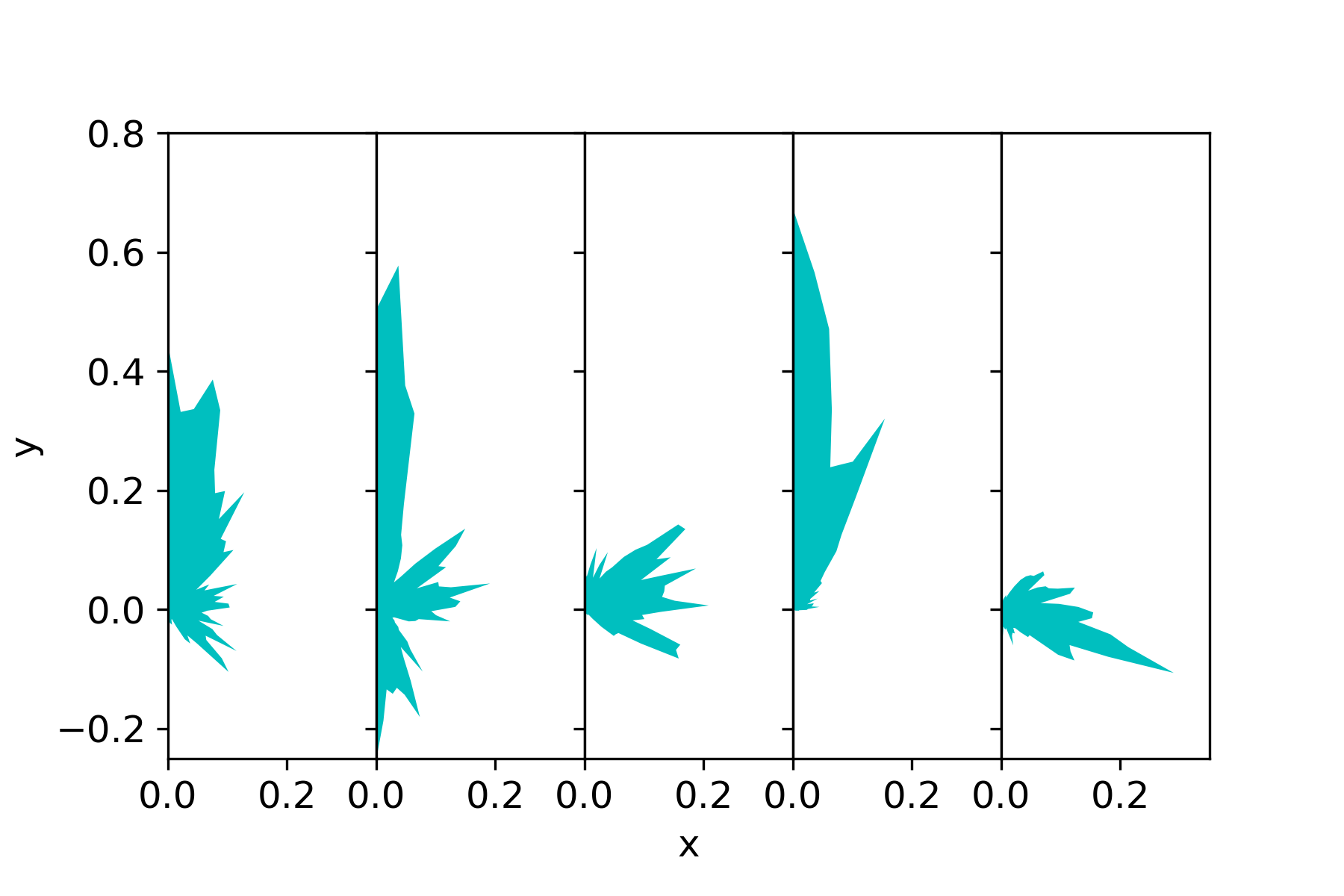}
    \caption{Examples of radially-parameterised shapes defined using $N=50$ coefficients and with a volume fixed to $V=0.01$.}
    \label{fig:Modified_Sphere - Examples}
\end{figure}

In this class the $N$ points defining the source boundary are given in polar coordinates as $(R_{\rm n}, \theta_{\rm n})$, where $\theta_n = \pi n/N$ and $R_{\rm n}$ is a product of the coefficients $b_k$ such that
\begin{equation}
    \label{eq:Perterbed Sphere - Definition}
    R_{\rm n} = \Pi_{k=0}^n b_k.
\end{equation}
For the case $b_k = 1$ for all $k$-values, the shape will approximately be a sphere (once the rotational symmetry is taken into account). To ensure the class contains well-behaved shapes we enforce that $b_k > 0$, so radial values are always positive. Also, since each radial value can be seen as the previous value multiplied by a factor $b_k$, by changing the range of values $b_k$ can take, we can control the smoothness of the shapes. This second point is important since we are refining the boundary of the shape, and so a large perimeter will require more cells, increasing the overall simulation time and memory required. We again will constrain the volume of the shapes to be a predetermined value $V$, therefore reducing the dimension of the shape space to $N-1$. This was done using the same method as the Legendre polynomial shapes, and is discussed in Appendix \ref{apx:Volume constraint}. Some examples of shapes contained in this class are shown in Figure \ref{fig:Modified_Sphere - Examples}.

\begin{figure}
    \centering
    \includegraphics{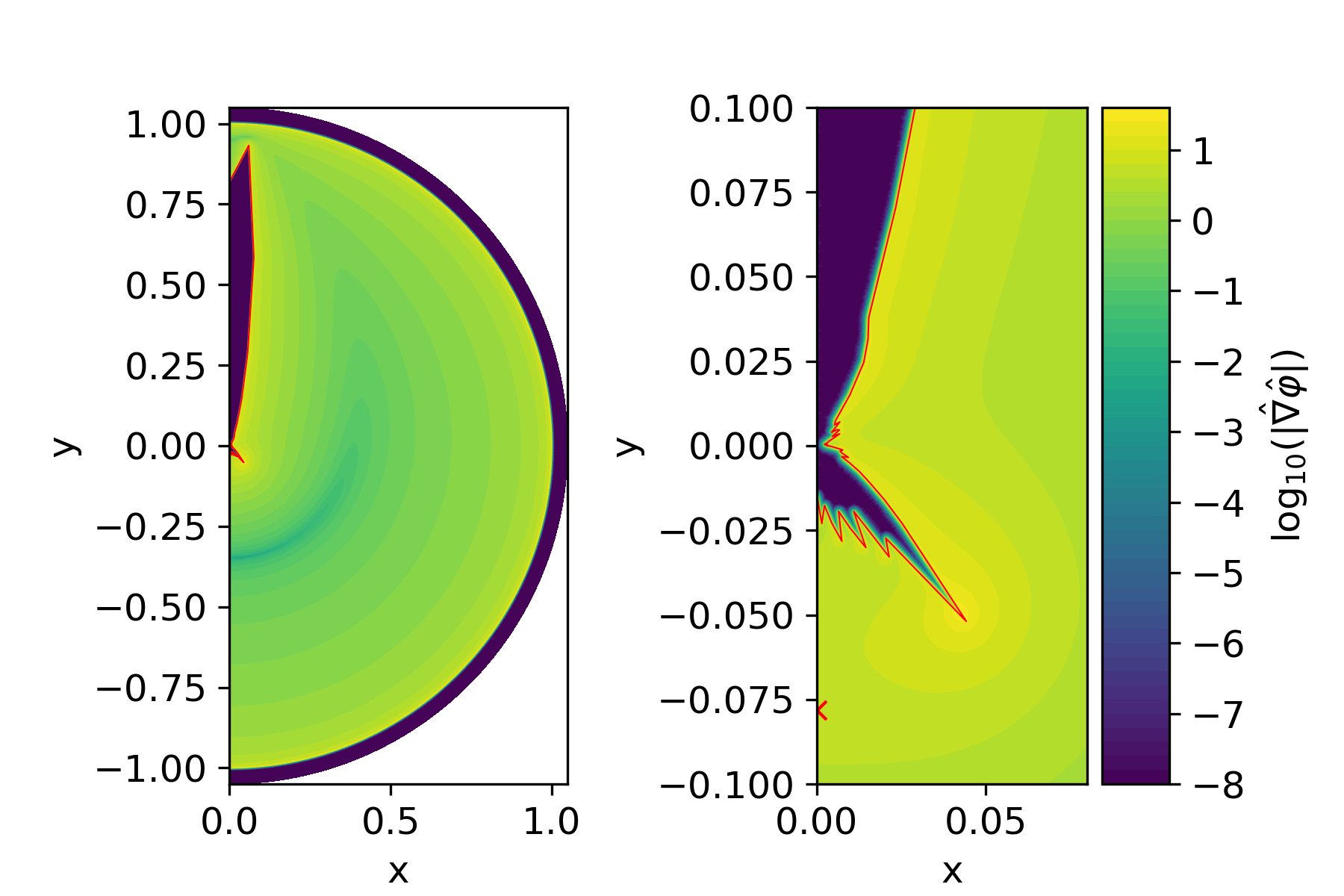}
    \caption{The chameleon force profile of a radially-parameterised source (boundary of which is indicated by the red line) inside a spherical vacuum chamber. The \textbf{left} plot contains the full solution while the \textbf{right} is zoomed in on the central region. The red cross on the $y$-axis below the source, at around y=-0.075, indicates the position where the force (at a distance from the source of $d=0.05$) is maximised. The value of this maximal force is $|\hat{\nabla} \hat{\varphi}|_{d} = 5.16$.}
    \label{fig:Mod_Sphere_GA_result_example}
\end{figure}

\begin{figure}
    \centering
    \includegraphics{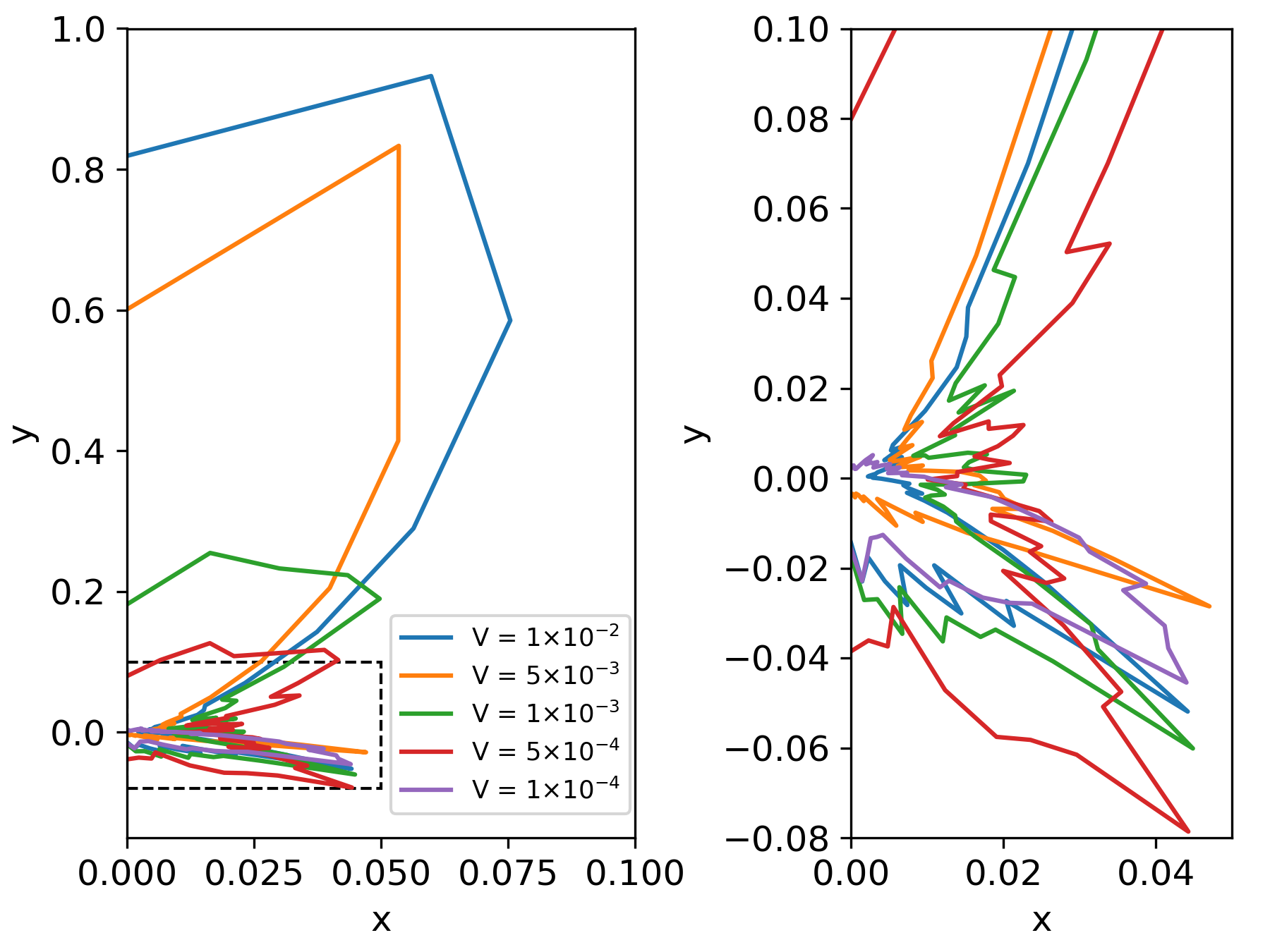}
    \caption{Shapes outputted by the GA when finding the optimal radially-parameterised shape with $N = 50$ coefficients that range between $0.5$ and $1.5$. Each line indicates a shape with a different volume, as shown in the legend, from $V=10^{-4}$ to $V=10^{-2}$. The \textbf{left} plot shows the full shapes, while the \textbf{right} is a zoomed in image of the area inside the dashed line.}
    \label{fig:Mod_Sphere - Compare R}
\end{figure}

We used the GA to optimise the force, a distance $d=0.05$ from the source, of a radially-parameterised shape defined by $N=50$ coefficients ranging between $0.5$ and $1.5$, and with a target volume of $V=0.01$. The shape outputted by the GA had a force of $|\hat{\nabla \hat{\varphi}}|_{d} = 5.16$ (which is slightly larger than the value obtained for the optimal Legendre polynomial shape), and was again located along the y-axis. The profile of the force generated by this optimised source shape is shown in Figure \ref{fig:Mod_Sphere_GA_result_example}, along with the position of the recorded value. Comparing this shape to the optimal Legendre polynomial shape shown in Figure \ref{fig:Legendre_GA_result_example}, we see that the GA has outputted a similar shape, with a large lobe around $\theta=0$ and a protruding spike (umbrella-like shape) in the lower region near $\theta = 3\pi/4$, despite being generated from two different shape parameterisations.
This similarity is not only present for the $V=0.01$ case. Comparing the GA results for radially-parameterised shapes of different volumes, we noticed similar shapes to those found when the GA was used with Legendre polynomial shapes for those same volumes, which can be seen by comparing Figures \ref{fig:Legendre - Compare R} and \ref{fig:Mod_Sphere - Compare R}. In section \ref{sec:Legendre polynomial shapes} we hypothesised that the umbrella-like feature might be the most important to maximising the force, and that the rest of the shape is just there to satisfy the volume constraint. These new results support this hypothesis since we see the same feature independent of the volume and for $V=10^{-4}$ we obtained only this feature without the upper lobe (note that it is not possible to obtain this shape with our Legendre polynomial parameterisation due to the property shown in equation (\ref{eq:Legendre R_max})).

\subsection{Is the umbrella the most impactful feature?}
\label{sec:Is the Umbrella the Impactful Feature?}

\begin{figure}[tbp]
    \centering
    \begin{subfigure}[b]{0.49\textwidth}
        \centering
         \includegraphics[width=\textwidth]{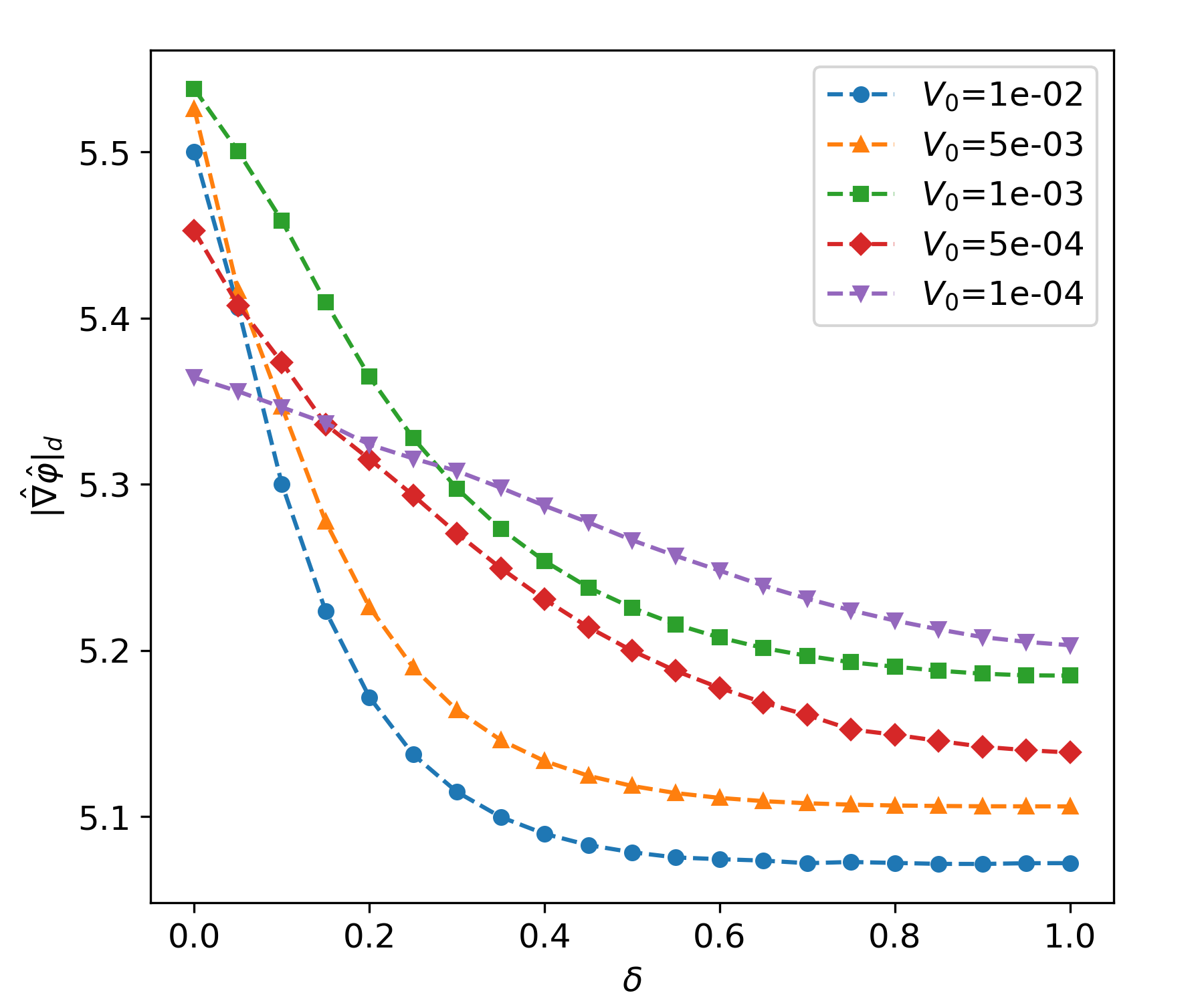}
         \caption{}
         \label{fig:cutoff - Force_vs_cutoff (Legendre)}
    \end{subfigure}
    \hfill
    \begin{subfigure}[b]{0.49\textwidth}
        \centering
         \includegraphics[width=\textwidth]{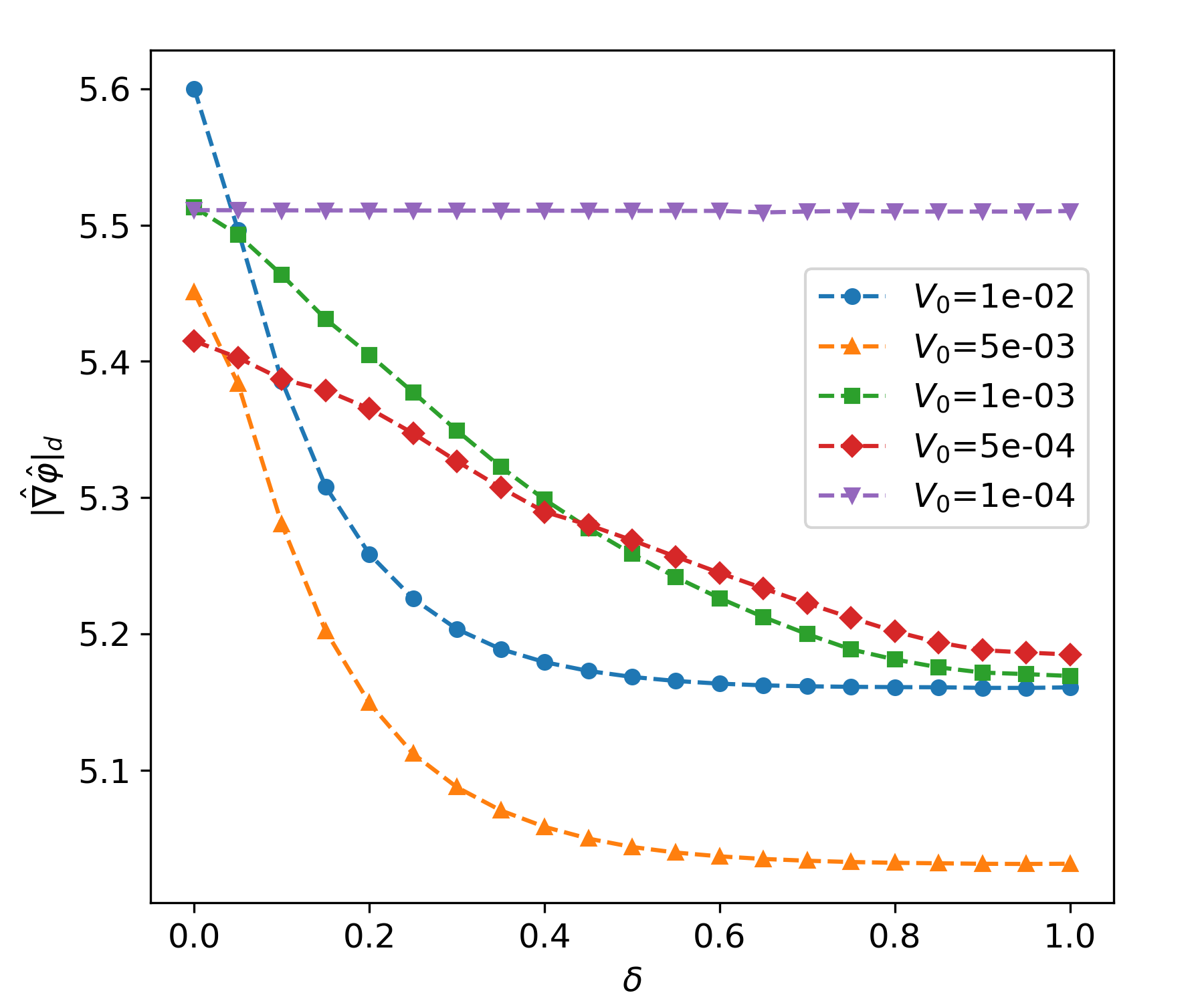}
         \caption{}
         \label{fig:cutoff - Force_vs_cutoff (radially_parameterised)}
    \end{subfigure}
    \hfill
    \caption{The force around Legendre polynomial (a) and radially-parameterised (b) shapes when parts of the upper lobe are removed. Each curve in plot (a) corresponds to one shape shown in Figure \ref{fig:Legendre - Compare R}, while in plot (b) the curves correspond to the shapes in Figure \ref{fig:Mod_Sphere - Compare R}. The volume of the unmodified shapes is shown in the legend.}
    \label{figs:cutoff - Force_vs_cutoff (both)}
\end{figure}

In sections \ref{sec:Legendre polynomial shapes} and \ref{sec:Radially-Parameterised Shapes} we found that for both the Legendre polynomial and radially-parameterised shapes, the GA appeared to converge to similar shapes. Furthermore, shapes obtained from different volume constraints consistently produced a similar feature, as seen in Figures \ref{fig:Legendre - Compare R} and \ref{fig:Mod_Sphere - Compare R}. As we impose rotational symmetry around the y-axis, the feature of interest is an umbrella shape. To test the importance of this umbrella feature we investigated how the measured force responds when other sections of the shapes, obtained in the previous two subsections, are removed. To do this we removed any part of the shape that lies above a cut-off height $y_c = \delta y_{\rm max}$, where $y_{\rm max}$ is the maximum $y$-value of the shape and $\delta \in [0,1]$. The resulting forces for the Legendre polynomial and radially-parameterised shapes are plotted in Figures \ref{fig:cutoff - Force_vs_cutoff (Legendre)} and \ref{fig:cutoff - Force_vs_cutoff (radially_parameterised)} respectively. In each case we see the position where the force is maximised is unchanged while its value increases as more of each shape is removed. This is in agreement with what was observed in section \ref{sec:Ellipsoid} (in Figure \ref{fig:Ellipsoid - Force_vs_volume}), in that above some critical volume the force increases with decreasing source volume. This further suggests that the umbrella shape, of this particular size, is the optimal shape we have been looking for. We will now test this hypothesis using a new shape class inspired by the umbrella shape in hopes to further improve our force. 

\begin{figure}
    \centering
    \begin{subfigure}[b]{0.49\textwidth}
        \centering
         \includegraphics[width=\textwidth]{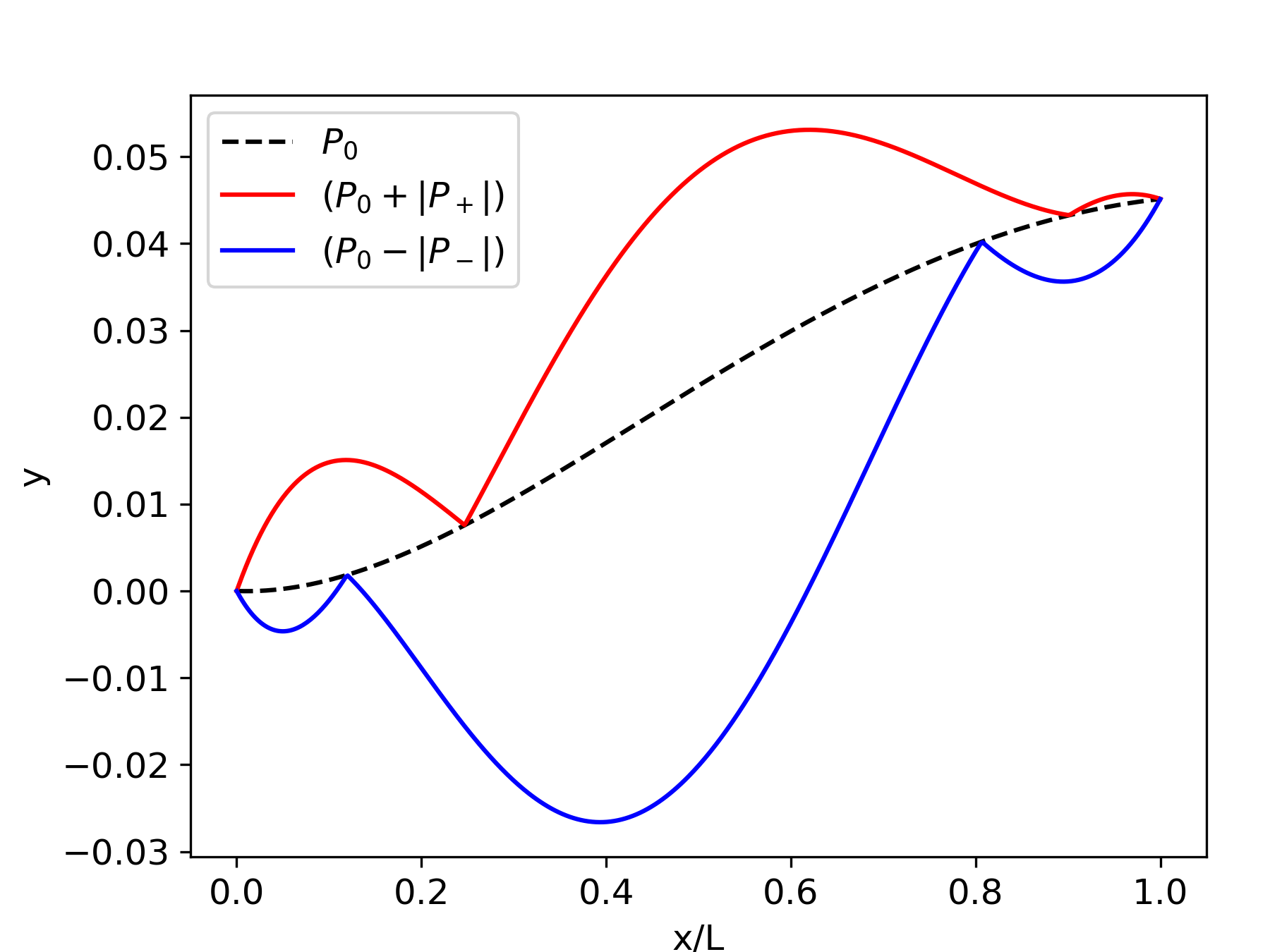}
         \caption{}
         \label{fig:Polynomial_Shapes - Shape examples (construction)}
    \end{subfigure}
    \hfill
    \begin{subfigure}[b]{0.49\textwidth}
        \centering
         \includegraphics[width=\textwidth]{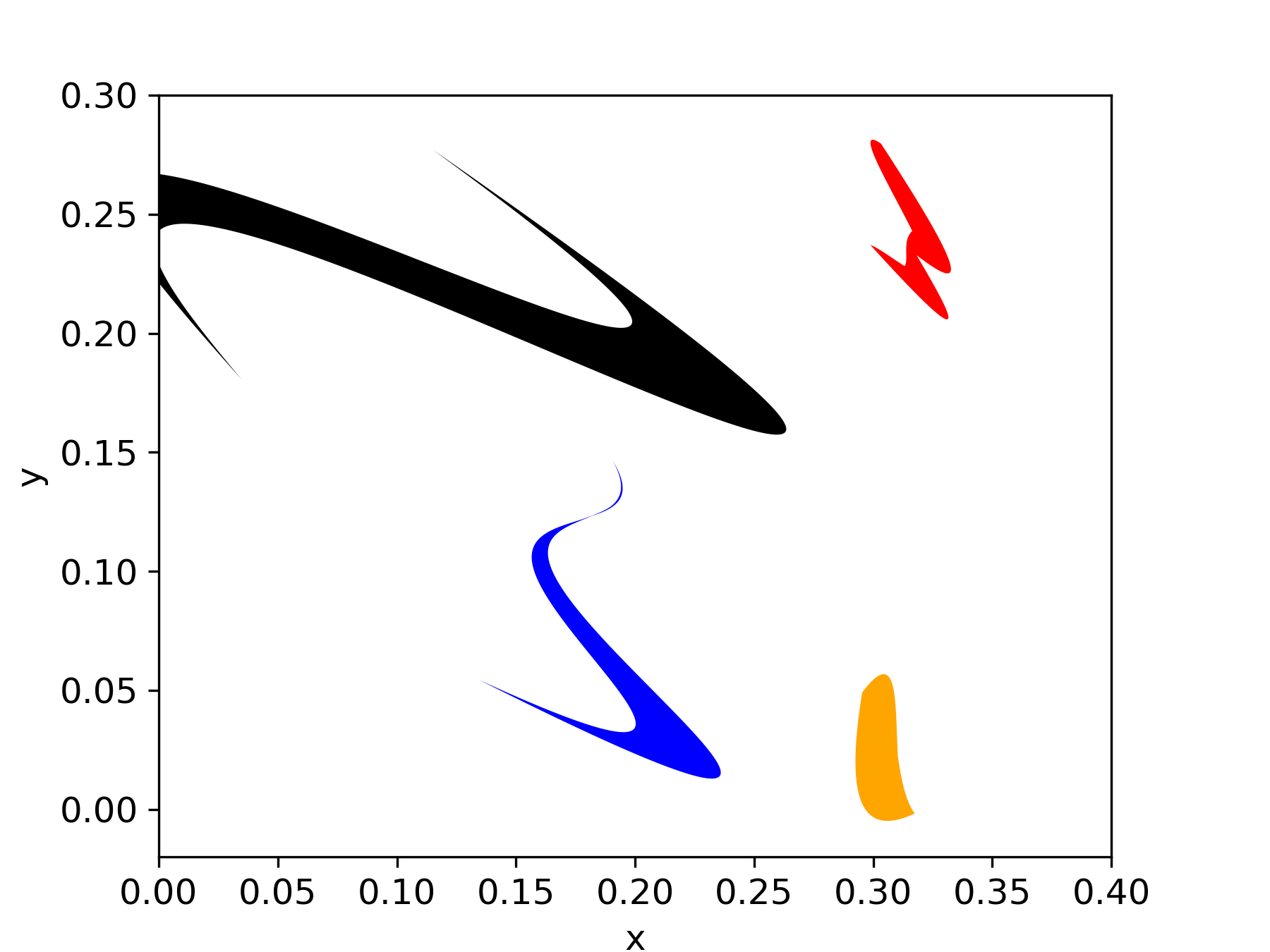}
         \caption{}
         \label{fig:Polynomial_Shapes - Shape examples (examples)}
    \end{subfigure}
    \hfill
    \caption{Plot (a) illustrates the construction of the polynomial class of shapes where $N=5$. The dashed line is the polynomial $P_0$, while the red and blue lines correspond to $(P_0 + |P_+|)$ and $(P_0 - |P_-|)$ respectively. The region enclosed defines the polynomial shape which is then translated and rotated according to the remaining shape coefficients. Plot (b) shows examples of shapes belonging to the polynomial class with coefficients bounded between $0$ and $0.3$.}
    \label{figs:Polynomial_Shapes - Examples}
\end{figure}

This new class consists of three polynomials (one of order $N$ and two of order $N+1$) defined over the range $x \in [0, L_{\rm u}]$, where $L_{\rm u}$ is a parameter of the shape. The first polynomial ($P_0$) acts as the baseline of the shape and is defined to have $N$ roots that are contained in the shape coefficients. The magnitudes of the second ($P_+$) and third ($P_-$) act as a displacement between $P_0$ and the boundary of the shape in the positive and negative $y$-directions respectively. These two polynomials are also defined by $N$ roots contained in the shape coefficients but also have an additional root at $x=L_{\rm u}$ so that the combined curves form a closed loop. For simple polynomials with not too many curves this will lead to umbrella-like shapes. However, this class still has the possibility of producing much more complicated shapes. For further generality the shape is then rotated around the origin anti-clockwise in the ($x$, $y$) plane by an angle $\theta$, and then translated by the vector $(x_0, y_0)$. Altogether, the number of parameters defining this class is $3N+4$. In this and future sections we used polynomials of order $5$ and $6$, so the number of coefficients is $19$, each ranging between $0$ and $0.3$. Unlike the previous shape classes, we place no constraints on the volume of the shape. We do, however, set a minimum positive value on the polynomials $P_+$ and $P_-$, to ensure that the curve does not intersect itself. A depiction of how shapes in this class are constructed, along with some examples, is shown in Figure \ref{figs:Polynomial_Shapes - Examples}. Among these examples one shows a case where the source has been separated into two disconnected pieces (shape filled in black). This occurred because to impose axis-symmetry we only consider parts of the shape for which $x>0$, therefore a sufficiently curvy shape can become two disconnected sources.

\begin{figure}
    \centering
    \includegraphics{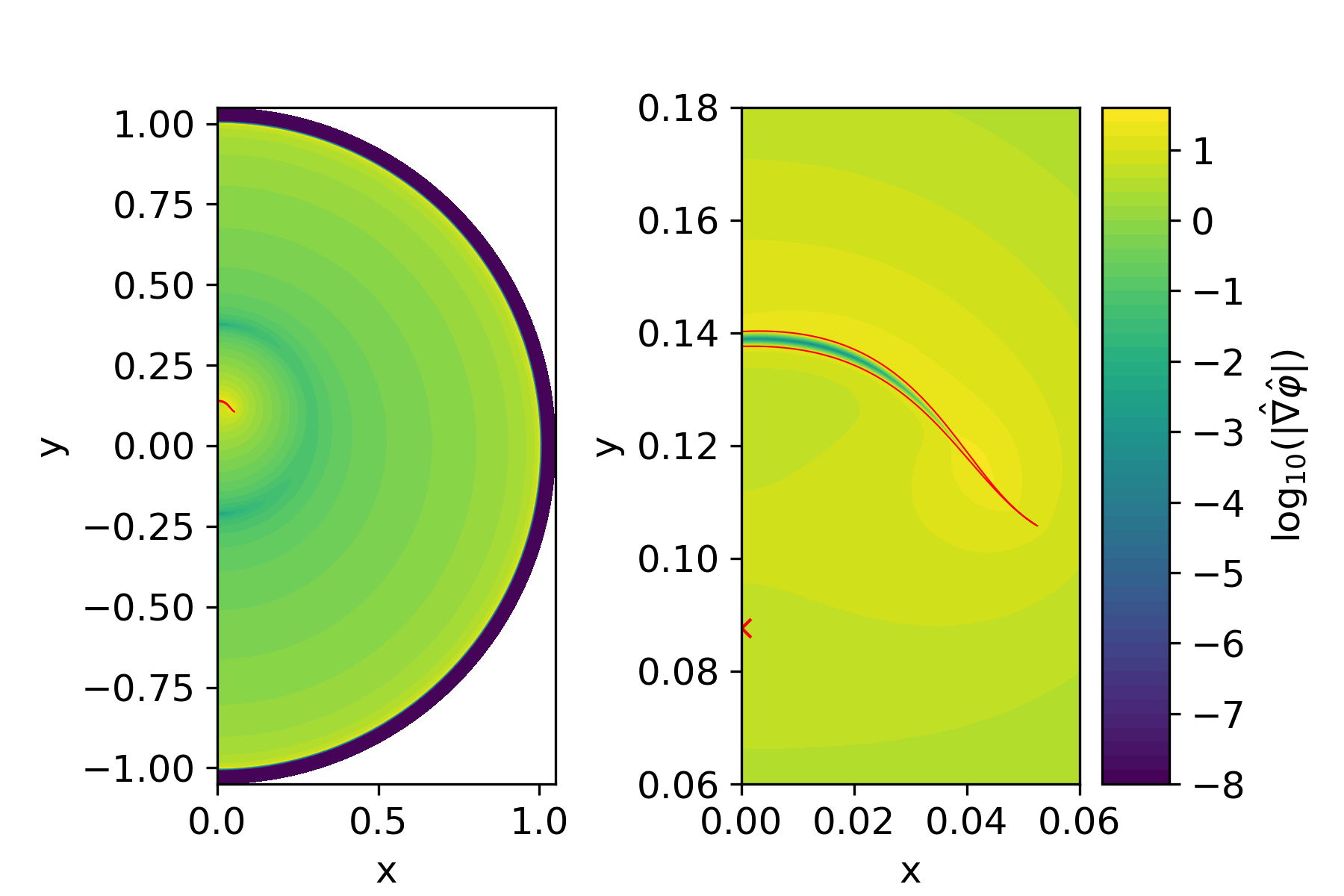}
    \caption{The chameleon force profile of a source whose shape belongs to the polynomial shape class where $N=5$ (the boundary of the shape is indicated by the red line), inside a spherical vacuum chamber. The \textbf{left} plot contains the full solution while the \textbf{right} is a zoomed in version showing the region immediately around the shape. The red cross on the $y$-axis below the source, at around $y=0.09$, indicates the position of where the force (at a distance from the source of $d=0.05$) is maximised.}
    \label{fig:Polynomial_GA_result_example}
\end{figure}

Using the GA to find the shape in this class that optimises the force we obtained the shape shown in Figure \ref{fig:Polynomial_GA_result_example}. We see the shape outputted by the GA is in agreement with our hypothesis that the umbrella was the optimal shape. Furthermore, thanks to the extra freedom provided by our parameterisation the outputted shape produces the best force value so far at $|\hat{\nabla} \hat{\varphi}|_{d} = 5.77$, and with a volume of $V = 1.26\times10^{-5}$. For scale this force is $2.45$ times larger than that generated by a sphere, located at the origin, of equal volume and using the same measuring distance. Furthermore, since the volume (and by extension mass) of the source is so small, this will lead to large ratios between the chameleon and gravitational forces.

\section{Effects of varying measuring distance}
\label{sec:Effects of Varying Measuring Distance}

\begin{figure}[tbp]
    \centering
    \begin{subfigure}[b]{0.49\textwidth}
        \centering
         \includegraphics[width=\textwidth]{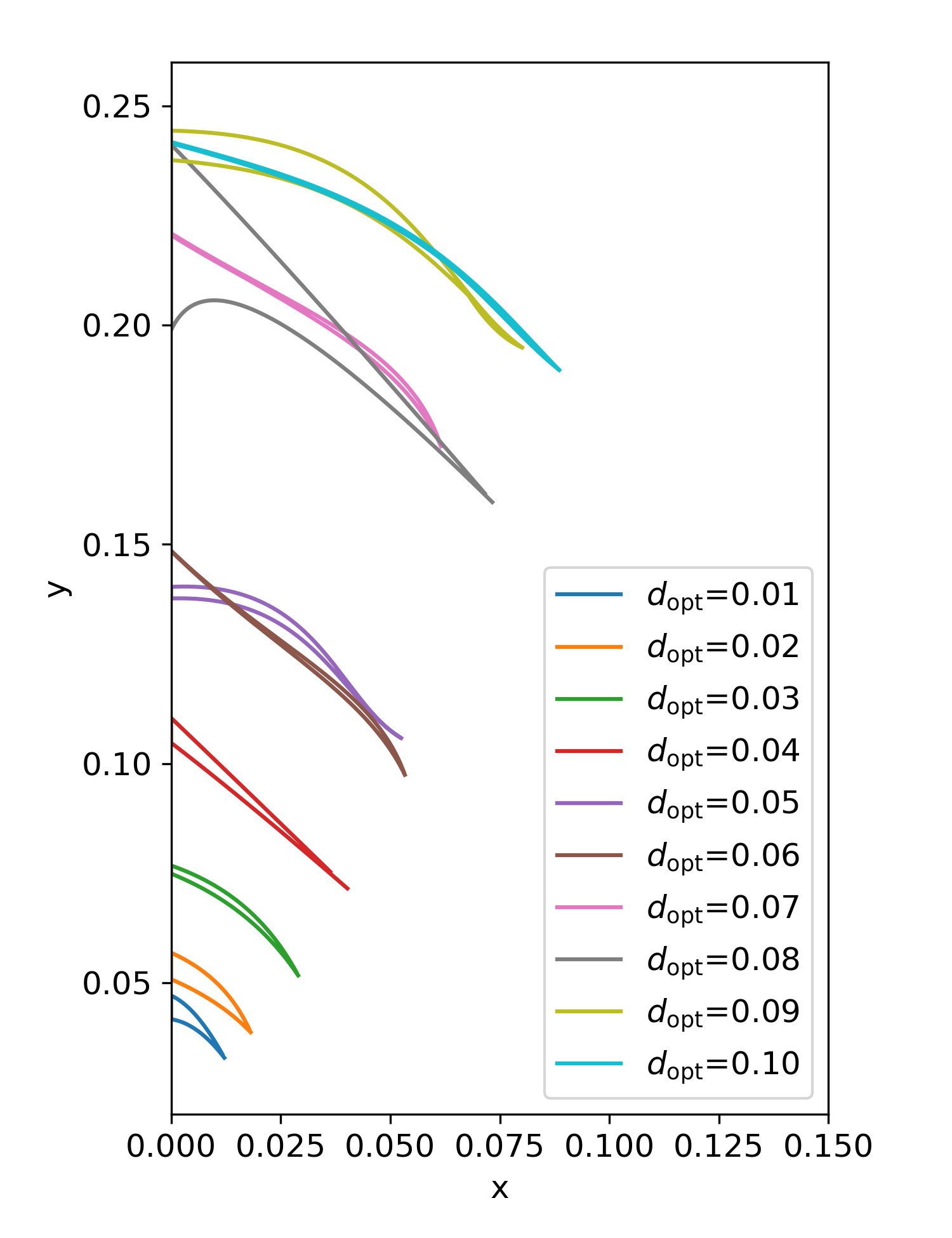}
         \caption{}
         \label{fig:Polynomial - GA results from varying d (shapes)}
    \end{subfigure}
    \hfill
    \begin{subfigure}[b]{0.49\textwidth}
        \centering
         \includegraphics[width=\textwidth]{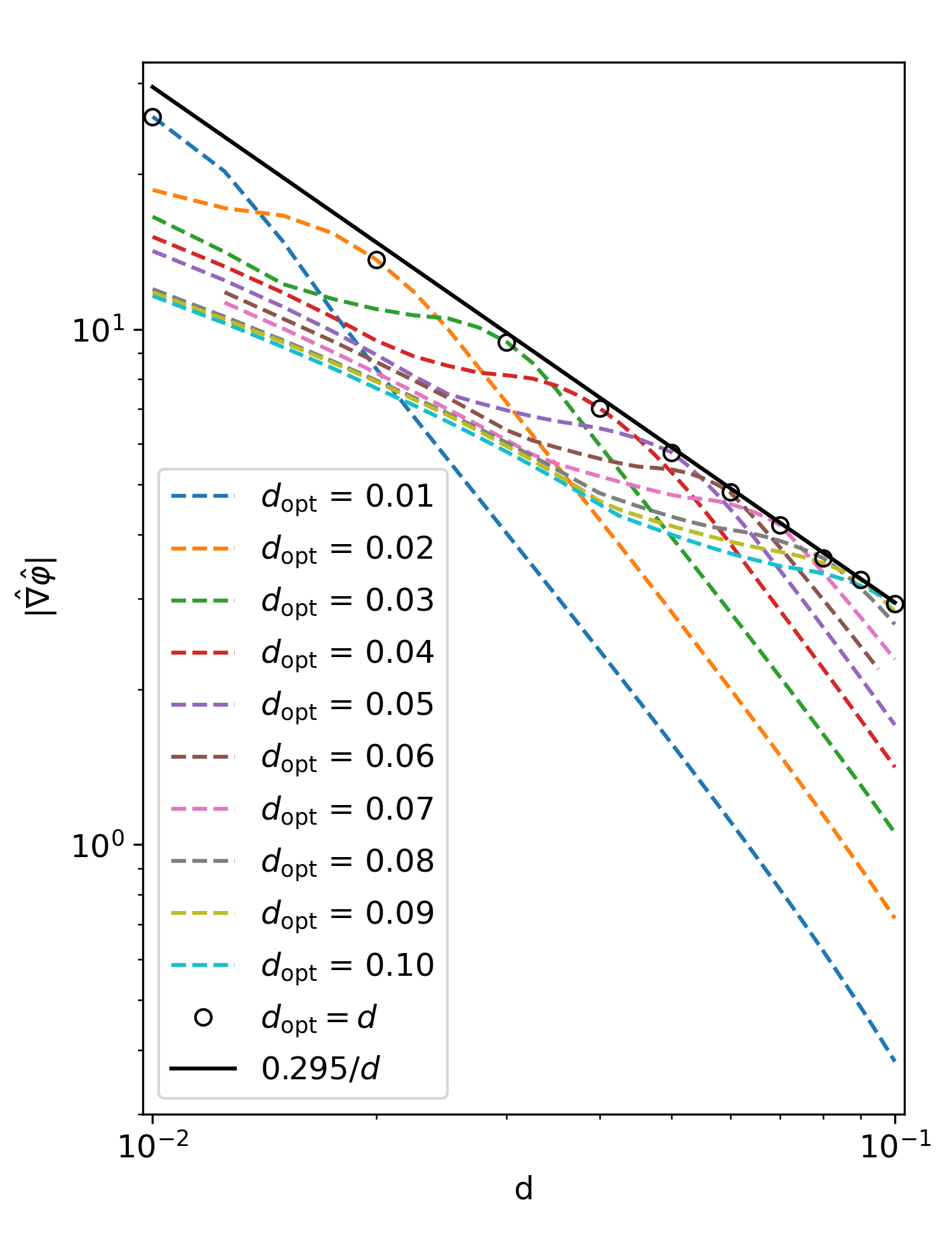}
         \caption{}
         \label{fig:Polynomial - GA results from varying d (plots)}
    \end{subfigure}
    \caption{Each line in plot (a) depicts the outline of a cross section of a polynomial shape (as described in section \ref{sec:Is the Umbrella the Impactful Feature?}) outputted by the GA when optimising the force at different measuring distances, $d_{\rm opt}$. For each of these shapes we then measured the maximum force at a range of distances, $d$, from the surface of the shape and plotted the resulting curves in plot (b). The black circles indicate when $d=d_{\rm opt}$, and the black line shows the curve $|\hat{\nabla}\hat{\varphi}|=0.295/d$.}
    \label{figs:Polynomial - GA results from varying d (both)}
\end{figure}

In this work so far the distance between where we wish to measure, and therefore maximise, the force and the surface of the source has been fixed at $d=0.05$. In this section we investigate how varying this distance, $d$, affects the optimal shape. We will use the polynomial shape class discussed in section \ref{sec:Is the Umbrella the Impactful Feature?} as it has provided the best results so far. The shapes shown in Figure \ref{fig:Polynomial - GA results from varying d (shapes)} are the results of the GA using this class, for varying values of $d$. We see that the umbrella shape remains the optimal shape found in each case, with the point of greatest force (at our fixed distance) being on the y-axis. It can also be seen that the vertical position and length of the shape does depend on the measuring distance, with both increasing with $d$. 
 
For each shape in Figure \ref{fig:Polynomial - GA results from varying d (shapes)} we measured the maximum force at a range of distances, the results of which are shown in Figure \ref{fig:Polynomial - GA results from varying d (plots)}. In this plot we have also indicated where on each curve the value $d$ is equal to the measuring distance the corresponding shape was optimised for, $d_{\rm opt}$. As expected each shape has a force that monotonically increases with decreasing distance; however, for $d > d_{\rm opt}$ the force decreases much faster than when $d < d_{\rm opt}$. Focusing on the points where $d = d_{\rm opt}$ we see that they approximately follow a $1/d$ relation, and the largest forces for any value of $d$ are from the shapes optimised at that specific measuring distance. This provides further evidence that the optimal shape does depend on the measuring distance in the experiment. This then means that when designing an experiment the minimum measuring distance (due to the experiment's design or from physical effects such as Casimir or Van der Waals forces) must be established first before the optimal source shape can be determined.

\section{Effects of varying n}
\label{sec:Effects of Varying n}

\begin{figure}[tbp]
    \centering
    \begin{subfigure}[b]{0.49\textwidth}
         \centering
         \includegraphics[width=\textwidth]{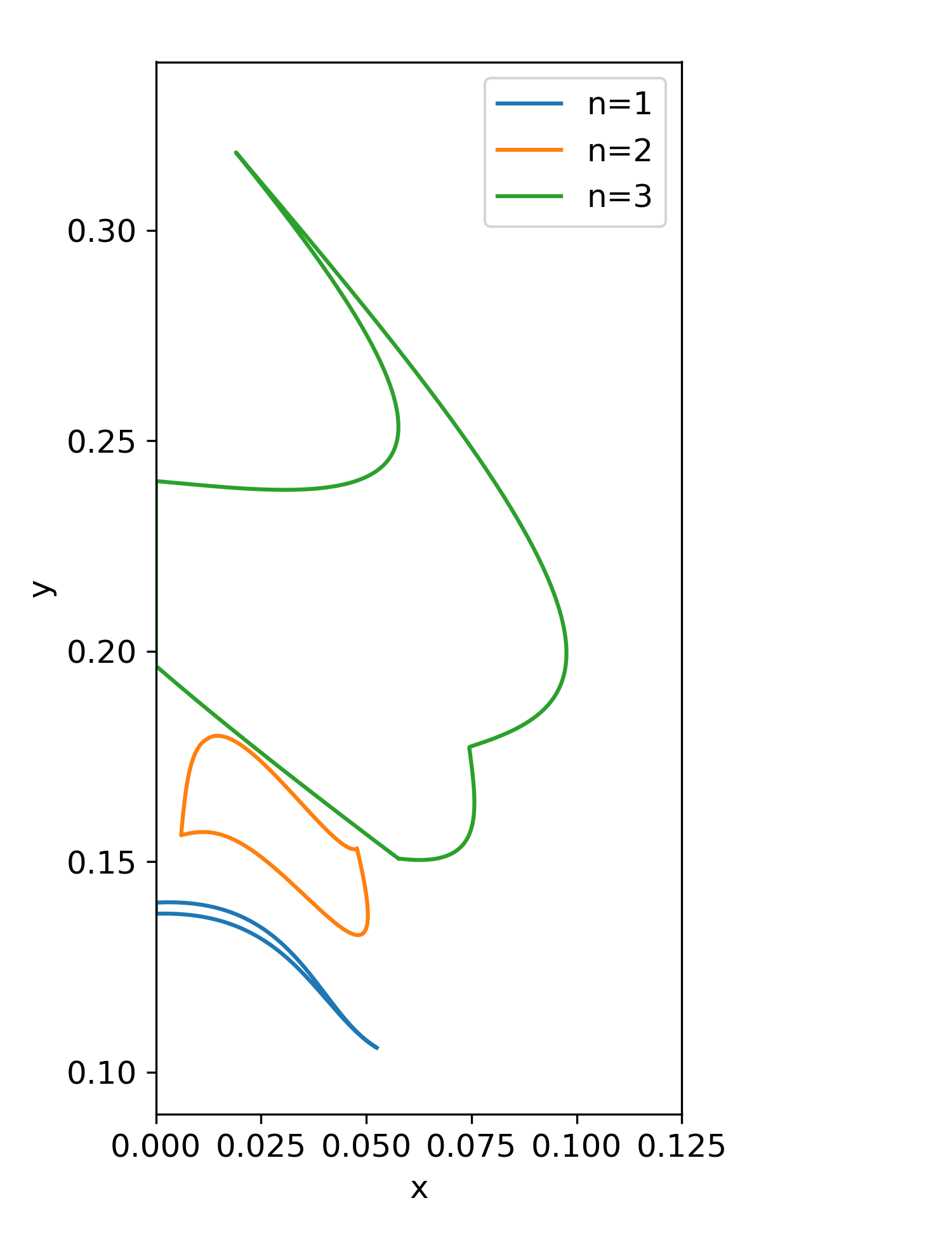}
         \caption{}
         \label{fig:Polynomial (vary n) - compare shapes}
    \end{subfigure}
        \hfill
        \begin{subfigure}[b]{0.49\textwidth}
        \centering
         \includegraphics[width=\textwidth]{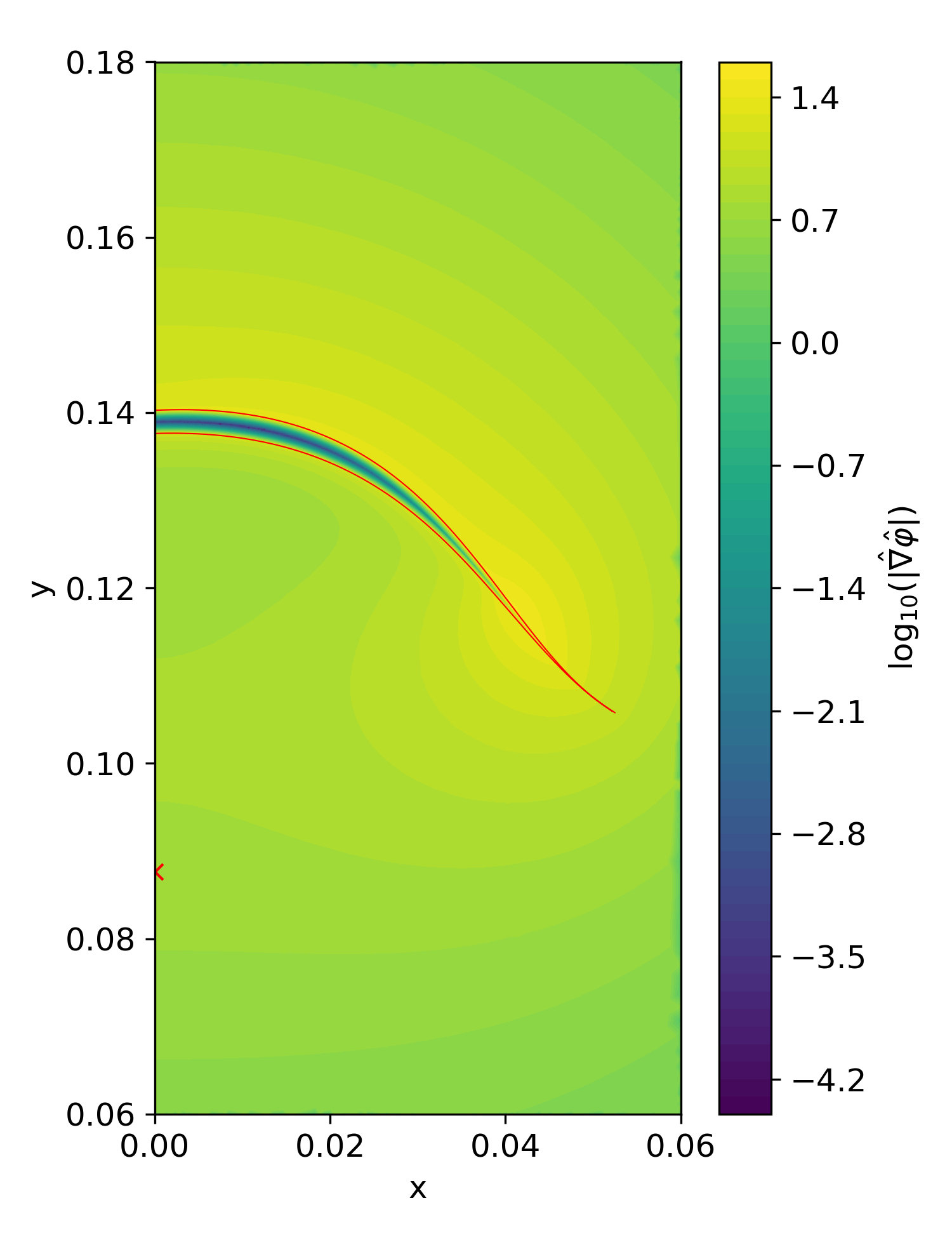}
         \caption{}
         \label{fig:Polynomial (vary n) - GA at n = 1.}
    \end{subfigure}
    \hfill
        \begin{subfigure}[b]{0.49\textwidth}
        \centering
         \includegraphics[width=\textwidth]{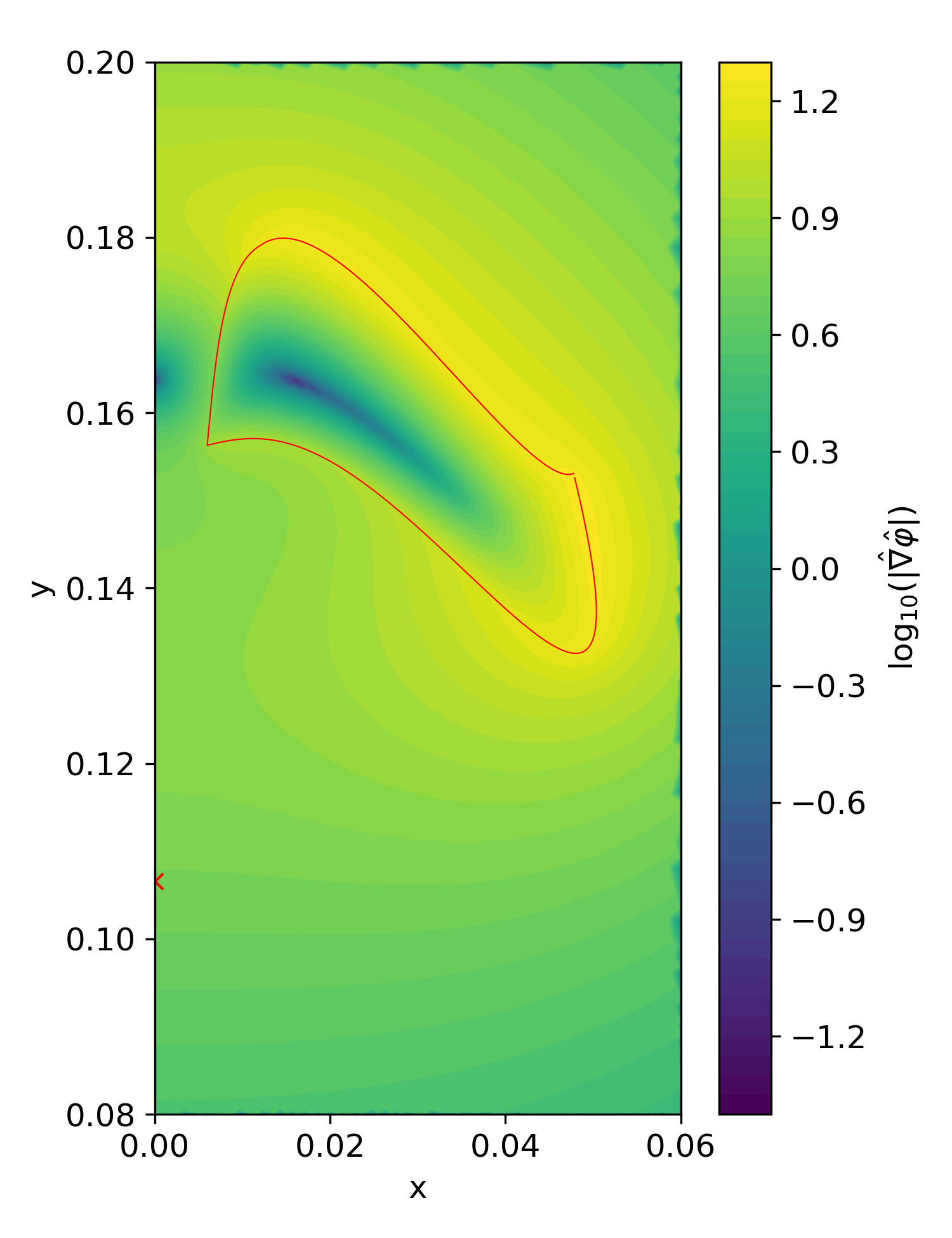}
         \caption{}
         \label{fig:Polynomial (vary n) - GA at n = 2.}
    \end{subfigure}
    \hfill
    \begin{subfigure}[b]{0.49\textwidth}
         \centering
         \includegraphics[width=\textwidth]{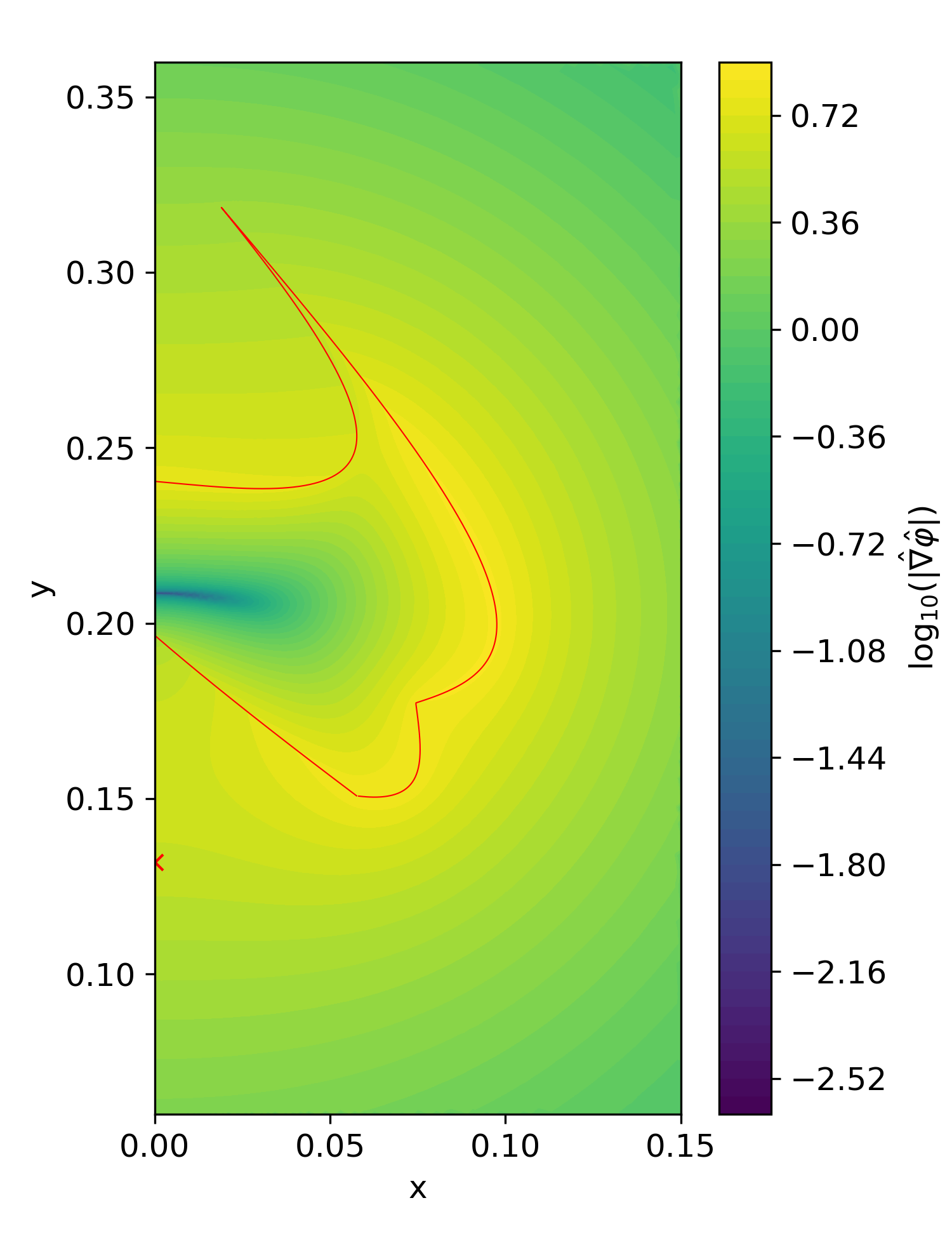}
         \caption{}
         \label{fig:Polynomial (vary n) - GA at n = 3.}
    \end{subfigure}
    \caption{Polynomial shapes outputted by the GA for fixed measuring distance $d=0.05$, but for varying $n$. Figure (a) shows the shapes compared to one another. Figures (b), (c), and (d) each show the chameleon force profiles sourced by the shapes in (a), for $n = 1$, $n=2$, and $n=3$ respectively. The red lines indicate the boundary of the shape.}
    \label{fig:Polynomial (vary n) - (all)}
\end{figure}

In the preceding sections we have explored the source shapes that could optimise the chameleon force measured in an experiment, for the potential shown in equation (\ref{eq:Chameleon - Potencial}) with $n=1$. However, this is not the only possible choice, and as more of the parameter space for $n=1$ is excluded, interest will increase in the $n>1$ chameleon models. To investigate how changing the value of $n$ would affect the optimal source shape, we used the GA to find the shape in the polynomial class (see section \ref{sec:Is the Umbrella the Impactful Feature?}) that optimises the force at a separation distance $d=0.05$ for $n=2$ and $n=3$. We note that $n$ can take some negative values that were not investigated in this work, but we expect the findings of this section to hold for these cases as well. The shapes outputted by the GA, and corresponding force profiles, are plotted in Figure \ref{fig:Polynomial (vary n) - compare shapes} along with the $n=1$ result discussed in section \ref{sec:Is the Umbrella the Impactful Feature?}, for comparison. We see that the umbrella again appears as the optimal shape in each case, but with an increasing thickness and vertical displacement with increasing $n$.

We see that the $n=2$ shape has a hole in it, and that the $n=3$ shape appears to be an umbrella with a `spike'. Looking at other shapes in the GA population for the $n=2$ simulations we find almost identical shapes that generate very similar forces as the optimal but with no hole. We therefore conclude that the existence of the hole contributes negligibly to the force. This is because the size of the hole is smaller than the field's Compton wavelength. Since the Compton wavelength is the inverse of the field's mass, as expressed in equation \ref{eq:Chameleon - Compton wavelength}, as $n$ increases so will the Compton wavelength and consequently smaller features will not affect the chameleon field profile and not contribute as significantly to the force. We also see this in the $n=3$ case where the shape outputted is consistent with the umbrella-like shape but with a spike. Looking at the force magnitude plotted in Figure \ref{fig:Polynomial (vary n) - GA at n = 3.} we see the spike contributes little to the force. The reason the GA outputted this shape can then be considered a consequence of an increasing Compton wavelength leading to a larger degeneracy in the shape to force mapping.

Now that we have determined that both the Compton wavelength and thickness of the shapes are increasing with $n$, we should determine whether the two effects are related by checking how much screening is occurring in each case. In this work we consider the interior of the source to be screened if the field has reached the value that minimises its effective potential in the source, $\hat{\phi}_{\rm c}$ (see equation (\ref{eq:Chameleon - Field minimum})), and the field gradient has a small magnitude, $|\hat{\nabla} \hat{\phi}| < 10^{-10}$. Under this definition we find that each shape shown in Figure \ref{fig:Polynomial (vary n) - (all)} is unscreened. However, measuring the field inside the sources, we found that the minimum field value was of the same order of magnitude as $\hat{\phi}_{\rm c}$. This combination of properties make intuitive sense as an unscreened source has all of its mass contributing to the field, and minimising the field's value inside the source allows for larger gradients around it. The reason then for the thickness of the shapes in Figure \ref{fig:Polynomial (vary n) - (all)} to increase with $n$ is because as the field's Compton wavelength increases, so does the thickness needed for the field to reach $\hat{\phi}_c$. This also helps to explain why the umbrella shape appears to be the optimal shape, since it maximises the amount of unscreened matter around the measuring position while also maintaining the optimal thickness.

\section{Conclusion}
\label{sec:Conclusion}

In this work we aimed to show that by exploiting the shape dependence of the chameleon field, we could improve experimental sensitivity through the use of an optimised source shape. We also aimed to determine what characteristics were necessary for such a shape. To accomplish this we used the software package SELCIE to numerically solve for the chameleon field profile of an axis-symmetric source mass inside a spherical vacuum chamber, and to measure the resulting force at a fixed distance from the source. The first class of shapes we investigated using this method were ellipsoids centred at the origin. We found that, for all volumes tested, as the ellipsoids deviate from the spherical limit the scalar force increases, with the largest forces being generated by disc-like ellipsoids. Additionally, we found that if the axis-ratio is fixed, then there is a critical volume which maximises the force, resulting in a preference for very small sources.

We investigated more complex shapes, with fixed volumes, using multiple shape parameterisations combined with a genetic algorithm, using the software package DEAP. We found that for each parameterisation used, the genetic algorithm would converge to shapes with a common feature, even when the volume of the shapes was varied, indicating a solution that is independent of the choice of parameterisation. After further investigation using a new shape parameterisation that did not constrain the volume of the source, we confirmed that the shapes that generate the largest force values are `umbrella'-like shapes, with the point where the force is maximised being located on the axis of rotation. The force generated by the optimum shape found (for $n=1$) was $2.45$ times larger than that of a sphere of equal volume located at the origin when the force is measured at the same distance to the source. 

We also found that the thickness of the umbrella is tuned so that it was both unscreened and had the field reach the value that minimises its effective potential inside the source. This means all the matter sources the field profile whilst also having as large a total field variation as possible, leading to high field gradients. We found that when changing the form of the field's potential, the thickness of the umbrella changed with the field's Compton wavelength, in such a way that this feature was present in each case tested. This characteristic is consistent with the critical volumes found for ellipsoids. When varying the measuring distance $d$, we found that the umbrella shape remains the optimal choice but its vertical position and length have a dependence on $d$. Furthermore, we found that the optimal force varies as $1/d$.

The umbrella shapes obtained in this work were designed to maximise the fifth force at a single point inside a spherical vacuum chamber. It is not necessarily a given, therefore, that they would be the best shape in an actual experiment, as experiments may use either moving particles, such as in atom interferometry, or an extended body as in torsion balance experiments. However, we believe that the property of the umbrella shapes, whereby the source is unscreened but the field reaches the value that minimises the effective potential inside the source, are generally necessary for optimising a fifth force experiment. Additionally, the method used in this work is general enough and sufficiently customisable that it can be tailored to specific experiments. One would only need to change the value outputted by the fitness function, obtained from the field solution, to whatever observable is used in that particular experiment. This is easily accomplished by extracting the required information from the field profile calculated using SELCIE, for example the total work done on a particle trajectory or scalar forces acting on an extended object. Any experimental constraints such as laser paths, particle trajectories, or different vacuum chamber shapes can also be enforced at the level of mesh generation using shape manipulation tools in SELCIE. These types of changes will likely break rotational symmetry and would therefore require solving the chameleon equations of motion using 3D meshes, which will greatly increase the simulation runtime and will therefore be left to future works.

\acknowledgments
Tamosiunas is supported by the Richard S. Morrison Fellowship. Briddon is supported by the University of Nottingham. Burrage and Briddon are supported by a Research Leadership Award from The Leverhulme Trust. Burrage and Moss are supported by the STFC under grant ST/T000732/1.

\appendix
\section{Genetic algorithm calibration}
\label{apx:Genetic Algorithm Calibration}

In this work we used DEAP \cite{DEAP_JMLR2012}, a software package designed to construct and perform genetic algorithm (GA) simulations. DEAP comes equipped with a wide range of options including choices for selection, crossover, and mutation methods, each with their own control parameters. Before using the GA to optimise our fitness function, as laid out in section \ref{sec:Genetic Algorithm}, we first tested the various possible configurations of the GA to find the most optimal one. This was accomplished using a test fitness function where the fitness value is the non-overlapping cross-sectional area of an input and target shape, both created using the Legendre polynomial parameterisation defined in section \ref{sec:Legendre polynomial shapes}. Therefore, the goal of the GA in these simulations is to minimise this fitness value and return coefficients that are as close to the target ones as possible.

\begin{figure}
    \centering
    \includegraphics{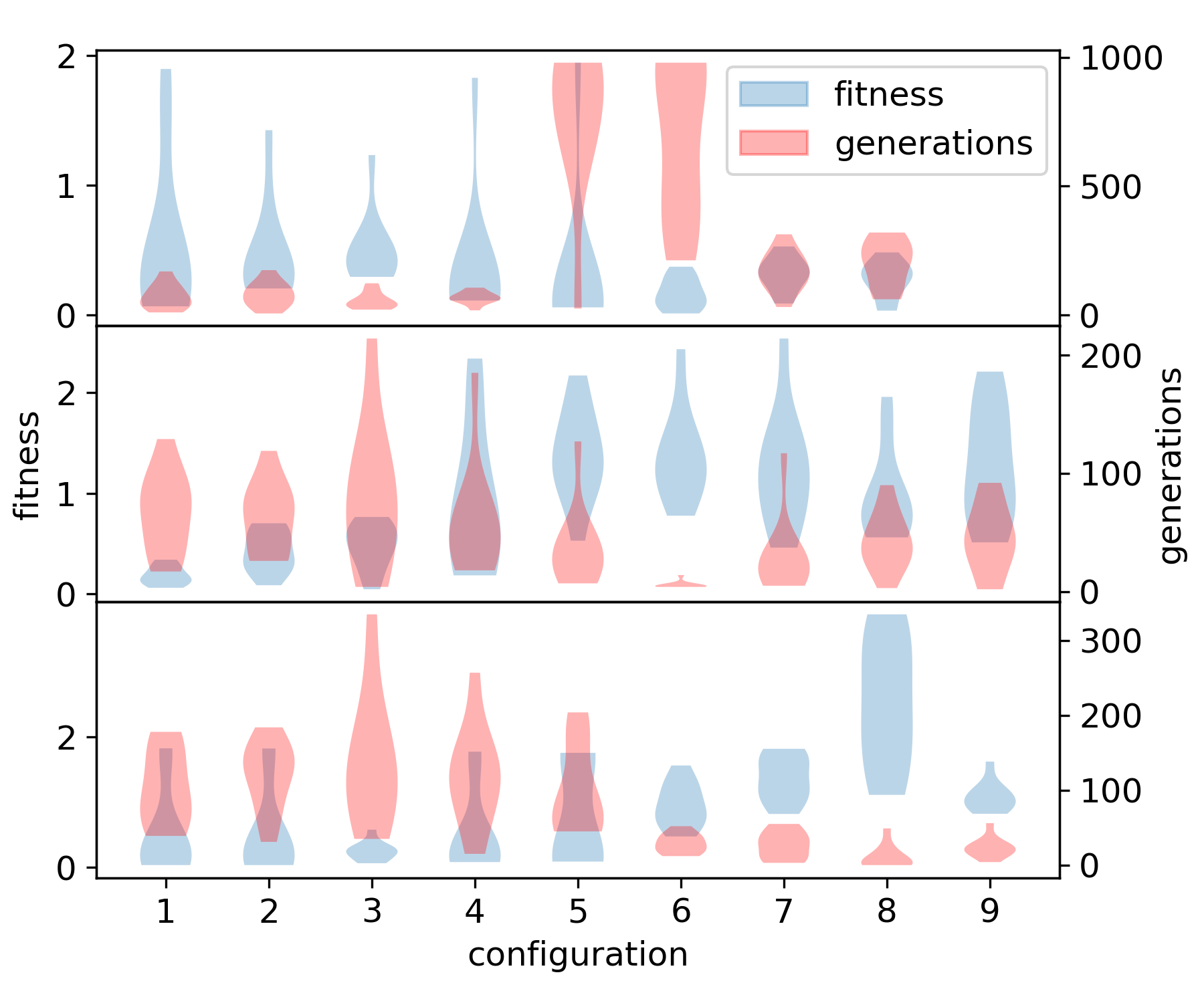}
    \caption{Violin plots showing the distribution of the number of generations taken to reach convergence and corresponding fitness values outputted by $10$ runs of the GA for a range of configurations, the details of which can be found at \url{https://deap.readthedocs.io/en/master/api/tools.html}. The fitness value is the non-overlapping cross-sectional area of the outputted shape and a target shape with coefficients $[0.5, 2, 0.9, 1.3]$, both of which are created using the Legendre polynomial parameterisation described in section \ref{sec:Legendre polynomial shapes}. In each plot the selection, crossover, and mutation stages not being varied are set as selTournament(tournsize=10), cxUniform(indpb=0.5), and mutGaussian(sigma=0.1), respectively. The number of individuals in each population was $100$. In the \textbf{top} plot the crossover method is varied with the configuration number, from left to right, corresponding to [cxOnePoint(), cxTwoPoint(), cxUniform(indpb=0.5), cxBlend(alpha=0.1), cxSimulatedBinary(eta=0.1), cxSimulatedBinary(eta=0.5), cxSimulatedBinaryBounded(eta=0.1), cxSimulatedBinaryBounded(eta=0.5)]. In the \textbf{middle} plot the mutation method is varied and the configurations are [mutGaussian(sigma=0.1), mutGaussian(sigma=0.3), mutGaussian(sigma=0.5), mutGaussian(sigma=0.7), mutGaussian(sigma=0.9), mutShuffleIndexes(), mutPolynomialBounded(eta=0.1), mutPolynomialBounded(eta=0.5), mutPolynomialBounded(eta=1.0)]. The \textbf{bottom} has the selection method being varied between [selTournament(tournsize=10), selTournament(tournsize=20), selTournament(tournsize=30), selTournament(tournsize=40), selTournament(tournsize=50), selNSGA2(), selSPEA2(), selRandom(), selBest()].}
    \label{fig (Apx):GA basic methods test}
\end{figure}

Due to the large number of possible configurations, we decided to treat our choice of selection, crossover and mutation methods as being independent from one another. This greatly reduces the number of configurations that require testing. For each of these configurations the GA was run $10$ times. The distribution of the number of generations needed to reach convergence and the corresponding fitness values outputted by the GA are both plotted in Figure \ref{fig (Apx):GA basic methods test} for a range of configurations. Additionally, we used this same test to evaluate other in-build algorithms available in DEAP, information on which can be found at \url{https://deap.readthedocs.io/en/master/api/algo.html}. Taking both the fitness distribution and number of generations until convergence into account we ultimately found the best choice of those tested was a custom algorithm that uses the methods: cxUniform(indpb=0.5) for crossover (configuration 3 in top plot of Figure \ref{fig (Apx):GA basic methods test}), mutGaussian(sigma=0.1) for mutation (configuration 1 in middle plot of Figure \ref{fig (Apx):GA basic methods test}), and selTournament(tournsize=10) for the selection method (configuration 1 in bottom plot of Figure \ref{fig (Apx):GA basic methods test}). Finally, we tested the effects of changing the input parameters of the chosen methods and found little effect on the distributions of the fitness values or number of generations until convergence.

\section{Rescaling shapes to constrain volume}
\label{apx:Volume constraint}
In our simulations the continuous space is replaced by a mesh consisting of discrete cells. For a 3D mesh these cells would consist of tetrahedrons with 4 vertices. In this case to measure the volume of our shape we would simply sum the volume of each individual cell that comprises it. For systems with an imposed symmetry, the system can be represented using a 2D mesh consisting of triangular cells made with three vertices. We therefore wish to know how to calculate the volume of the full 3D shape using its 2D intersection and knowledge of the symmetry.

For rotational symmetry we can imagine projecting each of the triangular cells comprising the shape around the axis of symmetry. For cells not touching the axis this will lead to triangular tori centred on the axis, while cells touching the axis will form cone segments. We find that the volume of one of these rotationally extended triangles is
\begin{equation}
    \label{eq(apx):rotationally extended triangles volume}
    V = \frac{2 \pi}{3} A(\delta_1 + \delta_2 + \delta_3),
\end{equation}
where $A$ is the area of the triangle and $\delta_i$, for $i \in \{1,2,3\}$, are the distances between the vertices of the triangle and the axis of rotation. To obtain the total volume of the shape, we iterate over every cell comprising the shape and calculate the volume contribution of each cell using equation \ref{eq(apx):rotationally extended triangles volume}. 

In this work, shapes are constructed by joining a series of points together by straight lines to form their boundary. These points are defined on a plane by the radial values $R_n$, for $n \in [0, N]$, and have an angular separation $\delta \theta = \pi/N$. Therefore the $nth$ triangle will be defined by the vertices at $(R_n, n\pi/N)$, $(R_{n+1}, (n+1)\pi/N)$, and the origin. Using equation (\ref{eq(apx):rotationally extended triangles volume}) and summing over all the triangles the total volume of the shape is
\begin{equation}
    \label{eq(apx) - shape volume}
    V = \frac{\pi}{3} \sin(\delta \theta) \sum_{n=0}^{N-1} R_{n} R_{n+1} [R_{n+1}\sin((n+1)\delta\theta) + R_{n}\sin(n\delta\theta)].
\end{equation}
By performing the rescaling $R_{i} \rightarrow \epsilon R_{i}$, we see from equation (\ref{eq(apx) - shape volume}) that the value transforms as $V \rightarrow \epsilon^3 V$. This means if we first calculate the volume using equation (\ref{eq(apx) - shape volume}) and define $\epsilon = (V/\Bar{V})^{1/3}$, then by rescaling the radial distances by $\epsilon$ this will act to fix the volume of the shape to be the target volume $V$.

\bibliographystyle{JHEP-custom}
\bibliography{References.bib}{}

\end{document}